\documentclass[12pt]{article}

\usepackage{scicite}
\usepackage{graphicx}
\usepackage{amsmath}
\usepackage{url}
\usepackage{times}
\usepackage [backend=biber,style=science]{biblatex}
\usepackage[utf8]{inputenc}
\usepackage[T1]{fontenc}
\bibliography{sigma}

\newenvironment{sciabstract}{
\begin{quote} \bf}
{\end{quote}}

\def\be{\begin{equation}}
\def\ee{\end{equation}}
\def\bea{\begin{eqnarray}}
\def\eea{\end{eqnarray}}

\def\reff#1{\ref{#1}}

\def\eqs#1#2{Eqs.~(\reff{#1}) and (\reff{#2})}
\def\fig#1{Fig.~\reff{#1}}

\def\figs#1#2{Figs.~\reff{#1} and \reff{#2}}
\def\sec#1{Sec.~\reff{#1}}
\def\tab#1{Table~\reff{#1}}

\def\cH{\mathcal{H}}

\def\cL{\mathcal{L}}

\def\Tr{\,\mathrm{Tr}}
\def\mev{\mathrm{Me\kern-0.1em V}}
\def\gev{\mathrm{Ge\kern-0.1em V}}
\def\tev{\mathrm{Te\kern-0.1em V}}
\def\fm{\mathrm{fm}}

\def\lqcd{\Lambda_\mathrm{QCD}}

\newcommand{\lsim}{ {\
\lower-1.2pt\vbox{\hbox{\rlap{$<$}\lower5pt\vbox{\hbox{$\sim$}}}}\ } }
\newcommand{\gsim}{ {\
\lower-1.2pt\vbox{\hbox{\rlap{$>$}\lower5pt\vbox{\hbox{$\sim$}}}}\ } }

\makeatletter
\DeclareRobustCommand{\text}{%
  \ifmmode\expandafter\text@\else\expandafter\mbox\fi}
\let\nfss@text\text
\def\text@#1{{\mathchoice
  {\textdef@\displaystyle\f@size{#1}}%
  {\textdef@\textstyle\f@size{#1}}%
  {\textdef@\textstyle\sf@size{#1}}%
  {\textdef@\textstyle \ssf@size{#1}}%
  \check@mathfonts
  }%
}
\def\textdef@#1#2#3{\hbox{{%
                    \everymath{#1}%
                    \let\f@size#2\selectfont
                    #3}}}
\makeatother

\def\somtitle{}

\title{\somtitle Ab-initio calculation of the proton and the neutron's scalar
  couplings for new physics searches}

\author
    {Sz. Borsanyi$^1$,
      Z.~Fodor$^{1,2,3}$,
C.~Hoelbling$^1$,
L.~Lellouch$^{4,\ast}$,\\
K.~K.~Szabo$^{1,3}$,
C.~Torrero$^{4,\dagger}$,
L.~Varnhorst$^{1,4}$\\
\\
\normalsize{$^1$Department of Physics, University of Wuppertal, D-42119 Wuppertal, Germany,}\\
\normalsize{$^2$ Institute for Theoretical Physics, E\"otv\"os University, H-1117 Budapest, Hungary}\\
\normalsize{$^3$J\"ulich Supercomputing Centre, Forschungszentrum J\"ulich, D-52428 J\"ulich, Germany}\\
\normalsize{$^4$Aix Marseille Univ, Universit\'e de Toulon, CNRS, CPT, Marseille, France}\\
\\
\normalsize{$^\ast$To whom correspondence should be addressed; e-mail: laurent.lellouch@univ-amu.fr}\\
\normalsize{$^\dagger$Present address: U-Hopper Srl, 38122 Trento, Italy}
}

\date{}

\begin{document}

\maketitle


\begin{sciabstract}
Many low-energy, particle-physics experiments seek to reveal new
fundamental physics by searching for very rare scattering events on
atomic nuclei.
The interpretation of their results requires quantifying the 
non-linear effects of the strong interaction on the spin-independent
couplings of this new physics to protons and neutrons.
Here we present a fully-control\-led, ab-initio calculation of these
couplings to the quarks within those constituents of nuclei.
We use lattice quantum chromodynamics computations for the four
lightest species of quarks and heavy-quark expansions for the remaining two.
We determine each of the six quark contributions with an accuracy
better than 15\%.
Our results are especially important for guiding and interpreting
experimental searches for our universe's dark matter.
\end{sciabstract}

\pagebreak


Many low-energy experiments, that search for new elementary particles
or interactions, are based on detecting very rare scattering events on
atomic nuclei. These include searches for weakly interacting massive
particles (WIMPs), that could constitute the elusive dark matter (DM)
of our universe. They also comprise the search for charged-lepton,
flavor-violating (cLFV) processes, whose observation would be a clear
sign of new, fundamental physics. Ambitious experiments are under
construction in both these areas. These may very well lead to the
discovery of new phenomena that are not described by the standard
model (SM) of particle physics. For a recent review of DM direct
detection experiments, see
\cite{Schumann:2019eaa}, and \cite{Mihara:2019kzi} for cLFV search
experiments.

On the theoretical side, the difficulty resides in making accurate
predictions for the rates expected in those experiments and in
interpreting their results. Atomic nuclei are composed of protons,
$p$, and neutrons, $n$, collectively known as nucleons, $N$. To
predict the expected rates and interpret the measurements, one must
quantify the interactions of the hypothetical new particles with
nucleons. The challenge is that nucleons are themselves complex,
nonlinear bound states of quarks, antiquarks and gluons. One must be
able to accurately describe these bound states to relate, what is
observed in experiment, to the fundamental parameters that describe
the interactions of the new particles with quarks and
antiquarks. These couplings are particularly important for the many
direct DM search experiments that attempt to detect WIMPs in the
spin-independent channel. They are also related to the low-energy
coupling of the Higgs boson, and of the ambient Higgs field, to
nucleons. In that sense, they are also associated with contributions
to the mass of nucleons. These couplings are known as nucleon
$\sigma$-terms.

We compute the $u$, $d$, $s$ and $c$-quark $\sigma$-terms of the
proton and neutron, using ab initio calculations based on quantum
chromodynamics (QCD), the fundamental theory of the strong
interaction. The Euclidean Lagrangian of QCD is $\cL = 1/(2g^2)\Tr\,
G_{\mu\nu}G_{\mu\nu} +\sum_q \bar
q[\gamma_\mu(\partial_\mu+iG_\mu)+m_q]q$, where $\gamma_\mu$ are the
Dirac matrices, $q$, the quark fields and index that runs over quark
flavors, and $g$ is the coupling constant. The $m_q$ are the quark
masses and $G_{\mu\nu}=\partial_\mu G_\nu-\partial_\nu
G_\nu+i[G_\mu,G_\nu]$, with $G_\mu$ the gluon field, a $3\times 3$
matrix in QCD. At the energies typical of quarks and gluons inside
hadrons, this theory exhibits highly nonlinear behavior. Thus we
introduce a hypercubic spacetime lattice on which the above Lagrangian
is discretized
\cite{Wilson:1974sk}, keeping quarks up to and including the charm.
The discretization puts quark variables on the lattice sites and gauge
variables on the links between neighboring sites. The discretized
theory is equivalent to a four-dimensional statistical physics system.
The Feynman path integral, which is used to define the quantum theory,
can thus be evaluated numerically using powerful, importance-sampling
methods \cite{Duane:1987de}. The calculation is performed for a number
of lattice spacings $a$ and spatial sizes $L$. The final results are
obtained after taking the limits $L\to\infty$ and $a\to 0$.

For the $b$ and $t$ quarks, we use a different strategy, based on a
sequence of heavy-quark effective field theories (HQET)
\cite{Shifman:1978zn}.  These are obtained from six-flavor QCD by
sequentially integrating out the most massive quark left in the
theory, starting from the top. As long as the masses of the heavy
quarks, $Q=t,b,\ldots$, integrated out are much larger than the typical
QCD scale, $\lqcd$, this can be done systematically in perturbation
theory, up to power corrections which begin at order
$(\lqcd/m_Q)^2$. The systematic error made in this approach is
determined by the powers of $\alpha_s(m_Q)$ and $(\lqcd/m_Q)$
appropriate for the order at which the calculation is performed, with
$Q$ corresponding to the lightest quark integrated out.

The nucleon $\sigma$-terms are conveniently pa\-ra\-me\-trized by the
dimensionless ratios, $f_{ud}^N=m_{ud}\langle N\vert\bar uu+\bar
dd\vert N\rangle/(2M_N^2)=\sigma_{\pi N}/M_N$ and $f_q^N=m_q\langle
N\vert \bar qq\vert
N\rangle/(2M_N^2)=\sigma_{qN}/M_N$ \cite{SOMsigma18}, where $N$ can be
either a $p$ or an $n$, the initial and final nucleon states have
identical momenta and are normalized relativistically, $q$ can be any
one of the six quark flavors, and $m_{ud}=(m_u+m_d)/2$. Note that in
the isospin limit, where $m_u=m_d$, $f_{ud}^n=f_{ud}^p$ and
$f_s^n=f_s^p$ and we will call these quantities $f_{ud}^N$ and
$f_s^N$, respectively.

$\sigma$-terms corresponding to the light $u$, $d$ and $s$ quarks have
a long history, as they concern one of the earliest low-energy
theorems established with current algebra \cite{Weinberg:1966kf}. This
has led to determinations of these $\sigma$-terms, based on $\pi
N$-scattering data and chiral perturbation theory
\cite{Gasser:1990ce,Pavan:2001wz,Hoferichter:2015dsa}.  There also
exist direct lattice calculations of the combined $u$ and $d$
$\sigma$-term \cite{
  Horsley:2011wr,Durr:2015dna,Bali:2016lvx,Yang:2015uis,
  Abdel-Rehim:2016won,Alexandrou:2019brg}, as well as some of $\sigma_{sN}$
\cite{
  Horsley:2011wr,Oksuzian:2012rzb,
  Freeman:2012ry,Junnarkar:2013ac,Gong:2013vja,Durr:2015dna,
  Yang:2015uis,Abdel-Rehim:2016won,
  Bali:2016lvx,Alexandrou:2019brg}, and two of the charm
\cite{Freeman:2012ry,Gong:2013vja,Alexandrou:2019brg}.
However, these calculations do not
exhibit full control over systematic errors or have too low
statistical precision.  Here we perform high-statistics lattice
calculations of the $\sigma$-terms for the four lightest quarks, in
which all relevant limits are taken with full control over
uncertainties. We use the Feynman-Hellmann theorem, which relates the
sigma terms to the partial derivative of $M_N$ with respect to the
corresponding quark mass. For $q=ud,s$, we proceed in two steps:
\begin{enumerate}

\item We calculate the logarithmic derivatives of $M_N$ with respect to
the $\pi^+$ and $K_\chi$ meson masses, $F_{P}^N\equiv (\partial\ln
M_N/\partial\ln M_P^2)$, with $P=\pi^+,K_\chi$. According to
leading-order SU(3) chiral perturbation theory ($\chi$PT),
$M_{\pi^+}^2\propto m_{ud}$ and $M_{K_\chi}^2\equiv
(M_{K^+}^2+M_{K^-}^2-M_{\pi^+}^2)/2\propto m_s$, so we expect these
derivatives to be close to the desired $\sigma$-terms.

\item We compute the Jacobian, $J$, for the coordinate transformation
$(\ln M_{\pi^+}^2,\ln M_{K_\chi}^2)\to (\ln m_{ud},\ln m_s)$ and
recover the $\sigma$-terms via $(f_{ud}^N,f_s^N)^T=J \cdot
(F_{\pi^+}^N$, $F_{K_\chi}^N)^T$.

\end{enumerate}
Thus, to obtain the $\sigma$-terms, we study $M_N$ as a function of
$M_{\pi^+}^2$, $M_{K_\chi}^2$ and $m_c$, as well as $M_{\pi^+}^2$ and
$M_{K_\chi}^2$ as functions of $m_{ud}$ and $m_s$. For that purpose,
we have used 33, $N_f=1+1+1+1$, 3HEX-smeared, clover-improved Wilson
ensembles, with a total of $18,300$ gauge configurations and $11$
staggered ensembles, with a total of $7280$ configurations, each
separated by 10 rational-hybrid-Monte-Carlo trajectories. The
staggered ensembles have $u,d,s$ and $c$ masses straddling their
physical values at three values of the bare gauge coupling,
$\beta=6/g^2$, corresponding to lattice spacings in the range $a=
0.06-0.10\,\fm$, and spatial extents around $6\,\fm$. The clover
ensembles have pion masses in the range of $195-420\,\mev$, the $s$
and $c$ masses straddling their physical values, four gauge coupling
corresponding to lattice spacings in the range $a=0.06-0.10\,\fm$ and
spatial extents up to $8\,\fm$. Details are given
in \cite{SOMsigma18}. In addition, as $M_N$ is well-known from
experiment \cite{Patrignani:2016xqp}, we use it to fix the lattice
spacing. In the $u,d,s$ sector, the physical mass point is defined by
setting $M_{\pi^+}^2$ and $M_{K_\chi}^2$ to their physical
values \cite{Aoki:2016frl}.

\begin{figure}
    \centering
    \includegraphics[width=0.49\columnwidth]{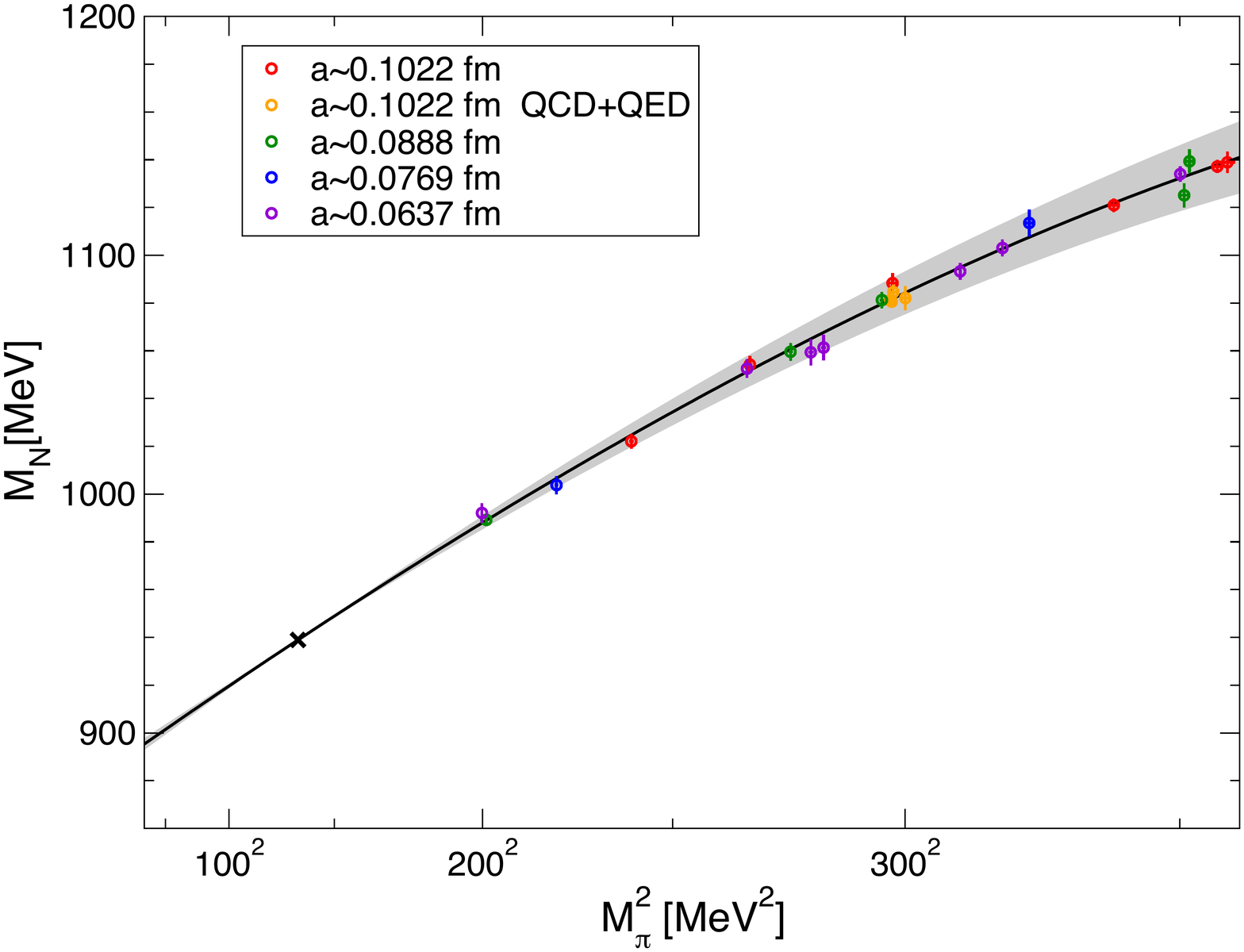}
    \includegraphics[width=0.49\columnwidth]{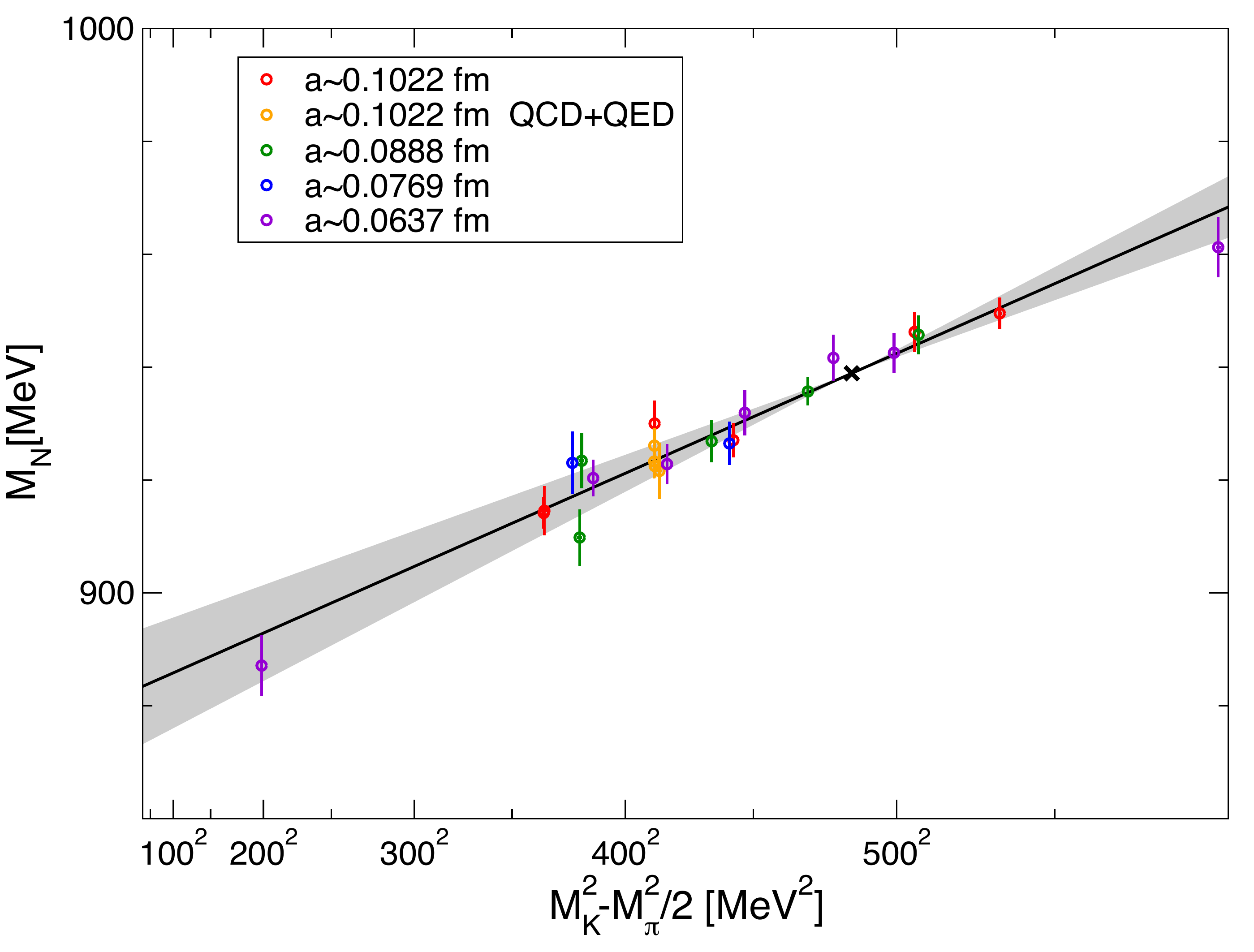}
    \caption
        {\label{fig:MNvsMpi2MKchi2}Example of the dependence of the
          nucleon mass on pion mass (left) and reduced-kaon mass
          (right) squared. The circles correspond to our simulation
          results for these masses at different values of the lattice
          spacing, indicated in the caption. The black cross in each
          plot corresponds to the physical value for these masses, as
          given in \cite{Patrignani:2016xqp,Aoki:2016frl}. The black
          curves with gray bands represent a typical fit to our
          results, with error bands. The values of
          $(F_{\pi^+}^{N},F_{K_\chi}^{N})$ obtained from the fit are
          given by the slope of the curves at the black cross. Note
          that all simulation points have been corrected, using the
          result of the fit, for the $M_{K_\chi}^2$ or $M_\pi^2$ and
          lattice spacing and volume dependencies that are not shown.
          All data
          points represent the mean $\pm$ SEM.}
\end{figure}

In \fig{fig:MNvsMpi2MKchi2}, we plot our clover determinations of
$M_N$ versus the values of $M_\pi^2$ and $M_{K_\chi}^2$, together with
the experimentally measured values for these quantities that define
the physical point. We fit these results for $M_N$ to various
polynomial, Pad\'e and $\chi$PT-motivated functions, of $M_{\pi^+}^2$
and $M_{K_\chi}^2$, that include discretization and finite-volume
corrections \cite{SOMsigma18}. From each fit we determine the
corresponding pair $(F_{\pi^+}^{N},F_{K_\chi}^{N})$ at the physical
point. A similar study of $M_{\pi^+}^2$ and $M_{K_\chi}^2$ versus
$m_{ud}$ and $m_s$ determines the four matrix elements of $J$ at the
physical point
\cite{SOMsigma18}. Combining these results, as described above, yields
the desired $f_{ud}^N$ and $f_s^N$. The statistical errors on these
results are calculated using 2000 bootstrap samples. To obtain
systematic errors, we consider 6144 different analyses each leading to
a result for $f_{ud}^N$ and $f_s^N$ and to a total goodness of fit
\cite{SOMsigma18}. These variations are chosen to probe the main
sources of systematic error. The results are then combined into
distributions whose means give our central values and whose widths
determine our systematic errors
\cite{Durr:2008zz,Borsanyi:2014jba,SOMsigma18} (see
\tab{tab:finalsigmaterms}).  Note that our analysis of the Jacobian
also provides a very precise determination of the mass ratio,
$m_s/m_{ud}=27.29(33)(8)$, where the errors are as in
\tab{tab:finalsigmaterms}. From $f_{ud}^N$ we further determine the
individual $u$ and $d$ contributions to the proton and neutron as in
\cite{Durr:2015dna} (see \tab{tab:finalsigmaterms}).

To determine $f_c^N$, we use 9 staggered ensembles at three $\beta$,
corresponding to lattice spacings $a=0.097-0.12\,\fm$.  For each $\beta$,
the ensembles feature three values of $m_c$ equal to $(0.75, 1, 1.25)$
times the physical value, $m_c^{(\phi)}$, with other quark masses held
at their physical values. We find that the most reliable way to obtain
$f_{c}^N$ is to consider the ratios of the nucleon correlator at
different $m_c$.  The exponential fall-off of this ratio gives the
difference of $M_N$ at these two values of $m_c$. From these ratios we
get the desired derivative either by a simple Taylor expansion or by a
heavy-quark-motivated expansion as detailed in
\cite{SOMsigma18} (see \tab{tab:finalsigmaterms}).

The $b$ and $t$ contributions are determined using the HQ expansions
discussed above. Here we use the next-to-next-to-next to leading order
result of \cite{Hill:2014yxa} to reduce the correction terms to
$O\left(\alpha_s(m_b)^4,(\lqcd/m_b)^2\right)$, leading to an
uncertainty from the HQ expansion of around $0.6\%$ \cite{SOMsigma18}.
The results are given in \tab{tab:finalsigmaterms}. With the same
approach, we can also compute $f_c^N$. But then the HQ expansion
uncertainties on the $c$, $b$ and $t$ contributions becomes
$O\left(\alpha_s(m_c)^4,(\lqcd/m_c)^2\right)$, i.e.  of order
$6\%$. We obtain $f_c^N|_\text{HQ}=0.0732(6)(7)$, where the errors are
as in \tab{tab:finalsigmaterms}. The difference to the full lattice
result is $f_c^N-f_c^N|_\text{HQ}=0.0002(45)(55)$, in good agreement
with the dimensional estimate of the possible HQ corrections, i.e. the
HQ expansion works as expected here.

\begin{figure}
 \centering
 \includegraphics[width=0.7\textwidth]{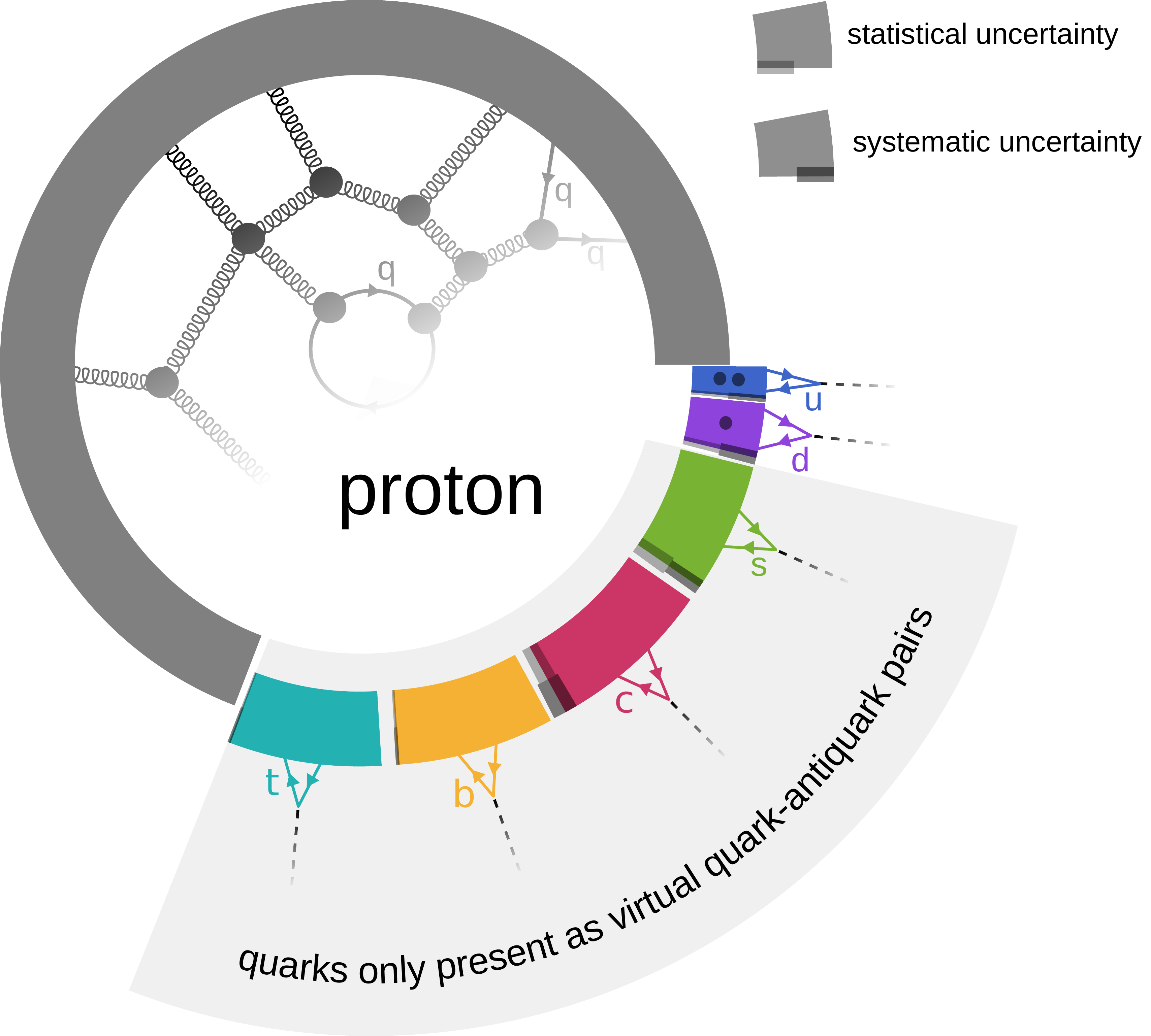}
 \caption{\label{fig_p_decomb}
 $\sigma$-terms of the proton shown as fractions of the total proton
 mass. Small black dots indicate the number of valence quarks per
 quark flavor. If the proton were an elementary fermion, the sum of
 these terms would be the mass of the proton. However, our calculation
 shows that this contribution is only $30\%$ (offset part) of the
 total, with less than $4\%$ coming from valence flavors.}
\end{figure}

In \fig{fig_p_decomb} we show our results for the $\sigma$-terms of
the proton, as fractions of the total proton mass. This representation
is renormalization scheme and scale independent, as described in
\cite{SOMsigma18}.  The $\sigma$-terms of the neutron differ from
those of the proton only in the small $u$ and $d$ contributions, at
the present level of precision. In addition, we provide a computer
code, based on our results, that allows to make predictions and to
interpret results of low-energy, experimental searches for new
fundamental physics. In particular, using it to sum all contributions,
we obtain the low-energy coupling of the Higgs to the nucleon,
$f_{hN}=0.3090(58)(61)$, in agreement with, but significantly more
precise than the result in \cite{Hoferichter:2017olk}.

Beyond its importance for direct dark matter searches and
charged-lepton flavor violation, our calculation also allows us to
complete the quantitative picture of how protons and neutrons acquire
mass, described, for instance, in
\cite{Durr:2008zz,Borsanyi:2014jba}. Instants after the Big Bang, the
universe is a hot gas of elementary particles. Quarks and gluons are
massless and interact through the strong interaction with a strength
that has been measured at the Large Hadron Collider (LHC). As the
universe expands and cools, it undergoes a
transition~\cite{Kajantie:1996mn,Csikor:1998eu} during which the Higgs
field acquires a non-zero expectation value. At that point, elementary
fermions get a mass through their interactions with the background
Higgs field. As the universe cools down further, in turn top, bottom
and charm quarks and antiquarks vanish through decay and annihilation,
and subsequently appear only as fleeting quantum fluctuations. The
universe can then be described by a theory of the strong interaction
in which these particles are completely absent, as long as their
fluctuations are subsumed into an increase in the strength of the
strong interaction. This is the context in which the calculations
of \cite{Durr:2008zz,Aoki:2006we} are performed. As the universe
continues to expand, it undergoes another transition, the QCD
crossover~\cite{Aoki:2006we}: the strongly interacting up, down,
strange quarks and antiquarks and gluons become confined within bound
states. In particular, protons and neutrons form out of up and down
quarks, with strange quarks, antiquarks and gluons contributing
through fluctuations. As shown in \cite{Durr:2008zz}, roughly $95\%$
of the mass of these two bound states comes about due to the energy
stored in the quantum fluctuations within them, while less than $5\%$
are induced by the up and down quark masses. The calculation performed
here reminds us that, within this $95\%$,
$\sum_{q=t,b,c}f_q^N=21.1(6)\%$ of $M_N$ is actually due to quantum
fluctuations of the massive $c$, $b$, $t$ quarks, with the remainder
being due to gluon and strange quark-antiquark fluctuations. And,
as \cite{Borsanyi:2014jba} confirmed, the permil difference between
neutron and proton mass arises from a subtle cancellation of
electromagnetic effects and effects due to the difference of up and
down quark masses.

\pagebreak

\subsection*{Acknowledgments}

We are indebted to S.~Collins, S.~D\"urr, J.~Lavalle, H.~Leutwyler and
E.~Nezri for informative discussions and correspondence. Computations
were performed on JU\-QU\-EEN and JURECA at Forschungszentrum
J\"ulich, on Turing at the Institute for Development and Resources in
Intensive Scientific Computing (IDRIS) in Orsay, on SuperMUC at
Leibniz Supercomputing Centre in M\"unchen, on HazelHen at the High
Performance Computing Center in Stuttgart.  This project was
supported, in part by the Excellence Initiative of Aix-Marseille
University - A*MIDEX (ANR-11-IDEX-0001-02), a French “Investissements
d’Avenir” program, through the Chaire d’Excellence program and the
OCEVU Laboratoire d’Excellence (ANR-11-LABX-0060), by the DFG Grant
SFB/TR55, by the Gauss Centre for Supercomputing e.V and by the
GENCI-IDRIS supercomputing Grant No. 52275.

\pagebreak

\begin{table}
  \centering
  \begin{tabular}{llll}
    \hline
    \hline
\multicolumn{2}{c}{Nucleon}
& \multicolumn{2}{c}{Individual $p$ and $n$}\\ \hline $f_{ud}^N$
    &0.0398(32)(44) & $f_{u}^p$ & 0.0142(12)(15)\\ $f_{s}^N$ &
    0.0577(46)(33) & $f_{d}^p$ & 0.0242(22)(30)\\ $f_{c}^N$ &
    0.0734(45)(55) & $f_{u}^n$ & 0.0117(11)(15)\\ $f_{b}^N$ &
    0.0702(7)(9) & $f_{d}^n$ & 0.0294(22)(30)\\ $f_{t}^N$ &
    0.0680(6)(7)
    &&\\ \hline \hline \end{tabular} \caption{\label{tab:finalsigmaterms}
    Final results for the $\sigma$-terms of the nucleons, in units of
    the corresponding nucleon mass. The statistical (SEM) and
    systematic uncertainties on the last digits are given in the first
    and second set of parentheses, respectively.}
\end{table}

\makeatletter
\renewcommand{\thefigure}{S\@arabic\c@figure}
\makeatother
\makeatletter
\renewcommand{\thetable}{S\@arabic\c@table}
\makeatother
\makeatletter
\renewcommand{\theequation}{S\@arabic\c@equation}
\makeatother

\newpage
\clearpage
\setcounter{table}{0}
\setcounter{figure}{0}
\setcounter{page}{1}
\setcounter{section}{0}
\setcounter{equation}{0}

\centerline{\Huge Supplementary Material}
\section{Nucleon scalar couplings}

\subsection{Feynman-Hellmann theorem}

Let $\vert N(\vec{p},r) \rangle$ be a relativistically-normalized,
free nucleon state with
three-momentum $\vec{p}$, helicity $r$ and mass $M_N$. It is an
eigenstate of the renormalized QCD Hamiltonian, $H_\mathrm{QCD}=\int
d^3x\;\cH_\mathrm{QCD}(x)$ with eigenvalue $E_{\vec{p}} =
\sqrt{M_N^2+\vec{p}^2}$. Then, applied to the momentum-independent
variation of $H_\mathrm{QCD}$ with respect to the renormalizaed mass
of quark $q$, $m_q^R$, the Feynman-Hellmann theorem
\cite{Guttinger1932,Pauli1933,Hellmann1933,Feynman:1939zza}
implies~\footnote{We thank Heiri Leutwyler for correspondence on the
  formulation of the Feynman-Hellmann theorem in quantum field
  theory.}
\begin{equation}
\frac{\partial M_N^2}{\partial m_q^R}=\frac12\sum_{r=1,2}\langle N(\vec{p},r) \vert\frac{\partial \cH_\mathrm{QCD}}{\partial m_q^R}(x)\vert N(\vec{p},r)\rangle_c\ ,
\end{equation}
where the subscript ``$c$'' indicates that we are referring to the connected
part of the matrix element. Now, according to \cite{Ji:1995sv}:
\begin{displaymath}
m_q^R\frac{\partial \cH_\mathrm{QCD}}{\partial m_q^R}(x)=m_q^R(\bar qq)^R(x)\ .
\end{displaymath}
so that the $\sigma$-terms are given by
\begin{eqnarray}
  \sigma^N_q & = & \frac{\partial M_N}{\partial\ln m_q^R}\nonumber\\
  & = & \frac1{4M_N}\sum_{r=1,2}\langle N(\vec{p},r) \vert m_q^R(\bar qq)^R(x)\vert N(\vec{p},r)\rangle_c\\
  & = & \langle N \vert m_q\bar qq\vert N\rangle/(2M_N)\nonumber\ ,
\end{eqnarray}
where, in the last line, we introduce a short-hand notation for the
previous expression, which takes into
account the fact that $m_q\bar qq$ is renormalization scale and scheme
dependent. From this one can define scalar, quark contents:
\begin{equation}
\label{eq:qc}
 f^N_q\equiv\frac{\partial\ln M_N}{\partial\ln
   m_q^R}=\frac{\sigma^N_q}{M_N}
\end{equation}
Note that both the $\sigma-$terms and the quark contents are
renormalization scheme and scale independent.

\subsection{Nucleon-Higgs coupling in the standard model}
In the standard model, the Higgs
field $\phi$ obtains a vacuum expectation value 
$$
\phi_0=\frac{v}{\sqrt{2}}=|\langle 0|\phi|0\rangle|
\ .$$
The quark masses, like all fundamental fermion masses, originate from
a Yukawa term $ g_f\bar f \phi f $ in the fundamental Lagrangian,
where $g_f$ is that fermion's Yukawa coupling and the fermion mass is
given by
\begin{equation}
\label{eq:mf}
m_f=g_f\phi_0
\ .
\end{equation}
Expanding $\phi=\phi_0+h$ around its vacuum expectation value, it is
obvious that the Yukawa term is also responsible for the fermion-Higgs
interaction, $g_f\bar f h f$ which, in principle, can be measured
independently. We can thus form the ratio
\begin{equation}
  f_f=\frac{g_f\phi_0}{m_f}
  \ ,
\end{equation}
which the standard model predicts to be one for fundamental
fermions. A similar ratio can be formed for fundamental gauge bosons
and it has been verified experimentally to be one for the $b$ and $t$
quarks, the $\tau$, $W$ and $Z$ \cite{ATLAS:2018doi}. We can slightly
rewrite this expression by noting that, according to (\ref{eq:mf}),
$$
\frac{\partial m_{f'}}{\partial\ln g_{f}}=g_{f}\delta_{ff'}\phi_0
\ ,$$
so that
\begin{equation}
\label{eq:fdef}
f_f^{f'}=\frac{\partial\ln m_{f'}}{\partial\ln g_{f}}
\end{equation}
is the fraction of the mass of fermion $f'$ that couples to the Higgs
field via the Yukawa coupling $g_{f}$ .

The advantage of the definition (\ref{eq:fdef}) is that it generalizes
naturally.  Returning to (\ref{eq:qc}), we note that the scalar quark
content is, in terms of fundamental parameters of the standard model:
$$
\label{eq:theone}
f^N_q=\frac{\partial\ln M_N}{\partial\ln g_q}
\ .$$
Thus, it can be interpreted as the mass fraction of the nucleon that
couples to the Higgs field via the Yukawa coupling $g_{q}$ of quark
$q$. Within the framework of the standard model, one may also
interpret $f^N_q$ as the fraction of the nucleon mass originating from
the Higgs field via the Yukawa coupling $g_{q}$. This definition is
scale and scheme independent, but it is not unique and we will
therefore not pursue it any further.

A large part of the nucleon mass does not couple to the Higgs field
via the $g_{q}$. The bulk of this ``rest'' originates from QCD
dynamics and there are standard techniques for decomposing this
contribution further \cite{Ji:1994av,Ji:1995sv,Yang:2018nqn}. Quantum
electrodynamics contributes at the permil level
\cite{Borsanyi:2014jba} and there are even smaller contributions due
to the weak interaction and the lepton sector, including its coupling
to the Higgs. Since the latter two do not take part in the strong
interaction, these contributions are suppressed by factors
$\alpha^2\sim 10^{-4}$ for the lepton sector and
$G_F\Lambda_\mathrm{QCD}^2\sim 10^{-5}$ for the weak dynamics
respectively. They can safely be ignored at the precision of our
current results.

Note that the sum over the scalar quark contents
of the nucleon,
\begin{equation}
  \label{eq:fhN}
f^N_h= \sum_q f^N_q
\ ,
\end{equation}
is an observable quantity. It denotes the strength of the coupling of
the Higgs to nucleons, in the limit of vanishing momentum
transfer. However, measuring it experimentally is a huge challenge,
since this requires isolating the highly-suppressed, Higgs exchange
contribution to the low-momentum-transfer scattering of a standard
model particle off a nucleon. Of course, in Higgs portal dark-matter
models, these interactions are the primary detection channel. In other
extensions of the standard model, scalar exchange interactions with
nucleons may require linear combinations of scalar quark contents that
differ from the one given in (\ref{eq:fhN}). For this reason, in
\sec{sect:tool} we provide a tool to compute arbitrary linear
combinations of quark contents from our results, taking the full
correlation matrix into account.

\section{Lattice details}

Different parts of the analysis rely on different gauge
configurations.  Some were obtained with a Wilson fermion action and
others with a staggered fermion one. These actions were chosen because
each has properties that are better suited to different aspects of
the calculation.

\subsection{Wilson gauge configurations}
\label{sec:wilsc}

In order to compute the dependence of the nucleon mass on the meson
masses used to interpolate to the physical $u$, $d$ and $s$ quark mass
point, we use $N_f=4\times 1$, 3HEX-smeared, clover-improved Wilson
fermions and a tree-level improved Symanzik gauge action (for details
see \cite{Borsanyi:2014jba}). Compared to \cite{Borsanyi:2014jba}, two
new ensembles were generated at the finest lattice spacing, with
strange quark masses significantly differing from the physical value,
so as to give us a larger lever arm in the strange quark mass
direction. The full list of these ensembles is provided in
\tab{ta:neu}.

\begin{table}
\begin{center}
\begin{tabular}{|c|c|c|c|c|c||c|c|c|}
\hline
$6/g^2$ & $e$ & $am_u$ & $am_d$ & $am_s$ & $L^3\times T$ & $m_\pi[\mathrm{MeV}]$ & $m_\pi L$ & \parbox{2cm}{\begin{center}$\times 1000$\\ trajectories\end{center}} \\
\hline
3.2 & 0 & -0.0686 & -0.0674 & -0.068 & $32^3\times 64$ & 413 & 6.9 & 1\\
3.2 & 0 & -0.0737 & -0.0723 & -0.058 & $32^3\times 64$ & 353 & 5.9 & 4\\
3.2 & 0 & -0.0733 & -0.0727 & -0.058 & $32^3\times 64$ & 356 & 5.8 & 1\\
3.2 & 0 & -0.0776 & -0.0764 & -0.05 & $32^3\times 64$ & 294 & 4.9 & 4\\
3.2 & 0 & -0.0805 & -0.0795 & -0.044 & $32^3\times 64$ & 238 & 4.0 & 12\\
3.2 & 0 & -0.0806 & -0.0794 & -0.033 & $32^3\times 64$ & 266 & 4.4 & 12\\
3.2 & 0 & -0.0686 & -0.0674 & -0.02 & $32^3\times 64$ & 488 & 8.1 & 4\\
3.2 & 0 & -0.0737 & -0.0723 & -0.025 & $32^3\times 64$ & 411 & 6.8 & 4\\
3.2 & 0 & -0.0776 & -0.0764 & -0.029 & $32^3\times 64$ & 336 & 5.6 & 4\\
3.2 & 0 & -0.077 & -0.0643 & -0.0297 & $32^3\times 64$ & 438 & 7.3 & 4\\
3.2 & 0 & -0.073 & -0.0629 & -0.0351 & $32^3\times 64$ & 469 & 7.8 & 4\\
3.2 & 0 & -0.077 & -0.0669 & -0.0391 & $32^3\times 64$ & 405 & 6.7 & 4\\
\hline
3.2 & 1.00& -0.0859 & -0.0792 & -0.0522 & $24^3\times 48$ & 298 & 3.7 & 5\\
3.2 & 1.00& -0.0859 & -0.0792 & -0.0522 & $32^3\times 64$ & 295 & 4.9 & 4\\
3.2 & 1.00& -0.0859 & -0.0792 & -0.0522 & $48^3\times 96$ & 295 & 7.3 & 4\\
3.2 & 1.00& -0.0859 & -0.0792 & -0.0522 & $80^3\times 64$ & 295 & 12.2 & 1\\
\hline
3.3 & 0 & -0.0486 & -0.0474 & -0.048 & $32^3\times 64$ & 422 & 6.1 & 1\\
3.3 & 0 & -0.0537 & -0.0523 & -0.038 & $32^3\times 64$ & 348 & 5.1 & 2\\
3.3 & 0 & -0.0535 & -0.0525 & -0.038 & $32^3\times 64$ & 349 & 5.0 & 2\\
3.3 & 0 & -0.0576 & -0.0564 & -0.03 & $32^3\times 64$ & 275 & 4.0 & 12\\
3.3 & 0 & -0.0576 & -0.0564 & -0.019 & $32^3\times 64$ & 293 & 4.2 & 12\\
3.3 & 0 & -0.0606 & -0.0594 & -0.024 & $48^3\times 64$ & 200 & 4.3 & 20\\
\hline
3.4 & 0 & -0.034 & -0.033 & -0.0335 & $32^3\times 64$ & 403 & 5.0 & 4\\
3.4 & 0 & -0.0385 & -0.0375 & -0.0245 & $32^3\times 64$ & 321 & 4.0 & 4\\
3.4 & 0 & -0.0423 & -0.0417 & -0.0165 & $48^3\times 64$ & 219 & 4.1 & 4\\
\hline
3.5 & 0 & -0.0218 & -0.0212 & -0.0215 & $32^3\times 64$ & 426 & 4.4 & 4\\
3.5 & 0 & -0.0254 & -0.0246 & -0.0145 & $48^3\times 64$ & 348 & 5.4 & 4\\
3.5 & 0 & -0.0268 & -0.0262 & -0.0115 & $48^3\times 64$ & 310 & 4.8 & 8\\
3.5 & 0 & -0.0269 & -0.0261 & -0.0031 & $48^3\times 64$ & 317 & 4.9 & 8\\
3.5 & 0 & -0.0285 & -0.0275 & -0.0085 & $48^3\times 64$ & 266 & 4.1 & 8\\
3.5 & 0 & -0.0302 & -0.0294 & -0.0049 & $64^3\times 96$ & 199 & 4.1 & 4\\
3.5 & 0 & -0.027 & -0.027 & -0.027 & $48^3\times 64$ & 280 & 4.4 & 3\\
3.5 & 0 & -0.028 & -0.028 & +0.009 & $48^3\times 64$ & 282 & 4.4 & 3.5\\
\hline
\end{tabular}
\end{center}
\caption{\label{ta:neu} List of 3HEX clover ensembles used to determine
  the meson-mass dependence of the proton mass.}
\end{table}

\subsection{Determination of hadron masses on Wilson ensembles}

We mainly use the pure QCD ($e=0$) ensembles. 
Pseudoscalar meson masses, $M_P$ are determined from a fit to multiple, zero-three-momentum correlators with a
common value of $M_P$:
\begin{equation}
C_{PP}(t)=\langle P(t)P^\dag(0)\rangle
\qquad\mbox{and}\qquad
C_{A_0P}(t)=\langle A_0(t) P^\dag(0)\rangle\ ,
\end{equation}
with $P=\bar q\gamma_5 q'$, $A_0=\bar q\gamma_0\gamma_5 q'$ and $q$
and $q'$ are distinct quark flavors chosen amongst $u$, $d$ and
$s$. We fit these correlation functions to
\begin{equation}
C_{PP}(t)=d\cosh(M_P(t-T/2))
\qquad\mbox{and}\qquad
C_{A_0P}(t)=f\sinh(M_P(t-T/2))\ ,
\end{equation}
where $T$ is the lattice temporal extent. Nucleon correlation functions are
those of \cite{Borsanyi:2014jba}. The corresponding masses were
obtained by fits to single, decaying exponentials.

Fit ranges are five lattice
spacings long and the initial time slices, $t_i$, are matched, on
different ensembles, to have a constant ratio between estimated
excited-state effects and the relative, statistical error on the nucleon
mass. Specifically, we take the smallest
$t_i>t_0-\ln(\epsilon_i)/\Delta M$, where $\epsilon_i$ is the
statistical error of the nucleon mass extracted with initial time
slice $t_i$, $\Delta M=500 \,\mev$\footnote{We checked that the
  exact offset value does not substantially change our results.} is an
estimate of the mass difference to the first excited nucleon state and
$t_0$ is a parameter which we vary in our analysis from $-1.4
\,\fm$ to $-1.25 \,\fm$ .

\begin{table}[h]
\begin{center}
  \begin{tabular}{ccc}
\hline
\hline
    $t_0$ & KS prob. $M_p$& KS prob. $M_n$\\
\hline
    -1.40 & 0.59 & 0.86\\
    -1.35 & 0.87 & 0.99\\
    -1.30 & 0.54 & 0.85\\
    -1.25 & 0.32 & 0.60\\
\hline
\hline
  \end{tabular}
\end{center}
  \caption{The Kolmogorov-Smirnov probability of the CDF of the proton
    and neutron mass fit qualities compared to a uniform distribution
    for various values of the offset $t_0$ (see text).
  \label{tab:platq}}
\end{table}

\tab{tab:platq} lists the Kolmogorov-Smirnov (KS) probabilities
from a comparison of the CDFs of the nucleon mass fit qualities to a
uniform distribution as detailed in \cite{Borsanyi:2014jba}. KS
probabilities for the multi-channel meson fits are around $0.2$
because of their highly correlated nature. Fitting channels
individually however gives fully compatible results and KS
probabilities that are in the same range as those for $M_p$ and
$M_n$. Therefore, we are confident that, by considering the full range of
offset values, $t_0$, from \tab{tab:platq} we obtain a
conservative estimate of remnant excited state effects.

For the extrapolation to the physical point, we used the ensembles
reported in \tab{ta:neu}. To get a better handle on the finite-volume
dependence of the nucleon mass, we added the four $e=1$, $\beta=3.2$
ensembles at $M_\pi\sim 290\,\mev$ that differ only in their volume
$L^3\times T$, with $L\in\{24,32,48,80\}$, compared to our set of pure QCD
ensembles. Since these ensembles were tuned to the isospin symmetric
point, we can use the neutron mass instead of the nucleon mass, the
connected-pseudoscalar-meson mass average, $(M_{uu}^2+M_{dd}^2)/2$,
instead of $M_\pi^2$ and $2M_{K^0}^2-M_{dd}^2$ instead of
$M_{K_\chi}^2$.

\subsection{Staggered gauge configurations and correlation functions}
\label{sec:stagc}

We use staggered ensembles near the physical point to extract the
dependence of the pseudoscalar meson masses, $M_P$, on the quark masses,
$m_q$. The full list of these ensembles
used in the analysis is
provided in \tab{ta:stg}.

\begin{table}
\begin{center}
\begin{tabular}{|c|c|c|c|c||c|c|}
\hline
$6/g^2$ & $am_{ud}$ & $am_s$ & $am_c$ & $L^3\times T$ & $m_\pi L$ & \parbox{2cm}{\begin{center}$\times 1000$\\ trajectories\end{center}} \\
\hline
3.84 & 0.00151556 & 0.0431935 & 0.511843 & $64^3\times 96$ & 4.1 & 5.1\\
3.84 & 0.00151556 & 0.04015 & .4757775 & $64^3\times 96$ & 4.1 & 3.25\\
3.84 & 0.00143 & .0431935 & .511843 & $64^3\times 96$ & 4.0 & 3.2\\
3.84 & 0.001455 & .04075 & .4828875 & $64^3\times 96$ & 4.1 & 15\\
3.84 & 0.001455 & .04075 & .4665875 & $64^3\times 96$ & 4.0 & 3.1\\
3.84 & 0.001455 & .03913 & .4636905 & $64^3\times 96$ & 4.0 & 5\\
\hline
3.92 & 0.001207 & 0.032 & 0.3792 & $80^3\times 128$ &  4.2 & 10\\
3.92 & 0.0012 & 0.0332856 & 0.39443436 & $80^3\times 128$ & 4.2 & 14.5\\
\hline
4.0126 & 0.000958973 & 0.0264999 & 0.314023 & $96^3\times 144$ & 4.1 & 1\\
4.0126 & 0.000977 & .0264999 & 0.314023 & $96^3\times 144$ & 4.2 & 10\\
 4.0126 & 0.001002 & 0.027318 & 0.323716 & $96^3\times 144$ & 4.2 & 2.7\\
\hline
\end{tabular}
\end{center}
\caption{\label{ta:stg} List of 4stout smeared staggered ensembles
  used to compute the quark-mass dependence of the pseudoscalar meson masses.}
\end{table}

The Goldstone-pion mass is close to the physical one in all of these
ensembles. The volumes are matched and $M_\pi L>4$, which implies that
the finite-volume corrections to the pion mass and decay constant are
below the permil level \cite{Colangelo:2005gd}. The kaon mass and decay constant receive even smaller
corrections. Since we do not determine any other observables from these
configurations, an infinite-volume extrapolation is not necessary in
this part of the analysis. Further details on the action used can be
found in \cite{Borsanyi:2013bia,Bellwied:2015lba}.

The dependence of the nucleon mass on the charm quark mass is obtained
directly from a set of nine 4-stout-smeared staggered ensembles
(``charm ensembles''), at three values of $\beta=3.75$, $3.7753$,
$3.84$. At each $\beta$ one ensemble, which we call the central
ensemble, was tuned to the physical point with a deviation of less
then 4 \% in $M^2_\pi/f^2_\pi, (2M^2_K - M_\pi^2)/f^2_\pi$ and the
bare charm quark mass was set to $11.85$ times the bare strange-quark
mass~\cite{Davies:2009ih}.  Two other ensembles were obtained from
each central ensemble by varying the bare charm-quark mass to $1.25$
and $0.75$ times the value on the central ensemble and leaving all
other parameters fixed. In each of these 9 ensembles, we have generated
64 configurations separated by $10$ trajectories each.

The sources and corresponding propagators are the standard ones
provided, for instance, by the MILC code \cite{MILCcode}.

\subsection{Determination of hadron masses on the staggered ensembles}
\label{sect:extract}

For the ensembles in \tab{ta:stg}, we obtain pseudoscalar
masses and decay constants from fits to the
pseudoscalar propagators starting at a fixed (in physical units)
$t_{\mathrm{min}}=1.9\,\mathrm{fm}$ or
$t_{\mathrm{min}}=2.3\,\mathrm{fm}$. Fit ranges are ten lattice
spacings long. In \tab{tab:platqs} we list the KS probabilities
that result from comparing the CDFs of the pion and kaon fit qualities
to a uniform random distribution.

\begin{table}[h]
\begin{center}
  \begin{tabular}{ccc}
\hline
\hline
    $t_{\mathrm{min}}[\fm]$ & KS prob. $M_\pi$& KS prob. $M_K$\\
\hline
    1.9 & 0.87 & 0.33\\
    2.3 & 0.14 & 0.77\\
\hline
\hline
  \end{tabular}
\end{center}
  \caption{Kolmogorov-Smirnov probability of the CDF of the pion
    and of the kaon mass fit qualities compared to a uniform distribution for
    various values of the offset $t_{\mathrm{min}}$.
  \label{tab:platqs}}
\end{table}

To extract masses from a staggered propagator, $c_t$, we either use a
standard staggered two state fit or, alternatively, we construct a
time-shifted propagator:
\begin{equation}
d_t(M)\equiv c_t+e^Mc_{t+1}
\label{eq:tsp}
\end{equation}
The time-shifted propagator is useful if there is a region in $t$ with
negligible backward contributions, $m(T/2-t)\gg 1$, and negligible excited
states except the parity partner of the nucleon.
Here $m$ is the ground state mass and $T$ the lattice
temporal extent. The staggered propagator in this region behaves as
\begin{equation}
c_t=e^{-mt}\left(
c_0+
(-1)^t
c_1 e^{-\Delta t}
\right)\ ,
\end{equation}
with the mass difference to the staggered parity partner, $\Delta$, and
the matrix elements $c_0$ and $c_1$ of the ground state and staggered
parity partner, respectively.

The contribution of the staggered parity partner to the time-shifted
propagator (\ref{eq:tsp}) can, in principle, be cancelled out by setting
$M=m+\Delta$:
\begin{equation}
d_t(m+\Delta)=
c_0 e^{-mt}(1+e^\Delta)\ .
\end{equation}
We can determine $M$ self-consistently by defining a local effective
mass
\begin{equation}
l_t(M)\equiv\ln(d_t(M)/d_{t+1}(M))\ ,
\end{equation}
as well as its average,
\begin{equation}
\bar{l}(M)=
\frac{1}{n}
\sum_{t=0}^{t_0+n-1}
l_t(m)
\end{equation}
in the signal region, $t_0\le t\le t_0+n$, and by minimizing
\begin{equation}
\sum_{t_1,t_2=0}^{t_0+n-1}
(l_{t_1}(M)-\bar{l}(M))
(C^{-1})_{t_1t_2}
(l_{t_2}(M)-\bar{l}(M))
\end{equation}
with respect to $M$, where $C$ is the correlation matrix corresponding
to $l_t(M)-\bar{l}(M)$. The resulting parameter
is $M_{\mathrm{opt}}$ and we call the corresponding propagator,
$d_t(M_{\mathrm{opt}})$, the optimal time-shifted propagator.

It is important to note that $M_{\mathrm{opt}}$ is just one number
per ensemble that fixes the relative contributions of $c_t$ and
$c_{t+1}$ in the time-shifted propagator. In particular, it does not directly
enter any further stages of analysis. After having determined
$M_{\mathrm{opt}}$, we proceed to extract the ground state mass from
$d_t(M_{\mathrm{opt}})$ in a standard fashion. In fact, the time-shifted propagator
$d_t(M)$ does have the correct asymptotic time behavior for any
constant $M$
and thus an inaccurate determination of $M_{\mathrm{opt}}$ does not
invalidate the ground-state-mass extraction from
$d_t(M_{\mathrm{opt}})$. If cancellation of the staggered parity
partner is not achieved to within the statistical accuracy of the correlator,
or additional excited states or backward contributions are present,
the consequent fit to determine the ground state mass from
$d_t(M_{\mathrm{opt}})$ will simply fail with a bad fit quality.

On our charm ensembles, we determine mass differences 
from ratios of optimal time-shifted propagators.  The plateau length
is always 8 lattice spacings and the plateau start is either
$t_{\mathrm{min}}=0.8\,\fm$ or
$t_{\mathrm{min}}=1.0\,\fm$.

\section{Computing mesonic $\sigma$-terms}

We determine the dependence of the nucleon mass on the pseudoscalar meson
masses from our 3HEX ensembles (\tab{ta:neu}). We define the isospin
symmetric physical point of QCD by $M_N=(M_p+M_n)/2=938.919\,\mev $
\cite{Patrignani:2016xqp} and $M_\pi=134.8\,\mev,
M_{K_\chi}=685.8\,\mev$ \cite{Aoki:2016frl}. In order to estimate the
dependence of our result on the range of the chiral expansion, we
imposed two different cuts on the maximal pion mass $M_\pi\le
360/420\,\mev$ entering our analysis.

\subsection{Fit forms}

The nucleon mass is extrapolated to the physical point with an ansatz
\begin{equation}
M_N=M_N^\phi\times
\prod_i
\left(
1+c_i(v_i-v_i^{(\phi)})
\right)
^{t_i}\ ,
\end{equation}
where the $c_i$ are the fit parameters, $v_i$ the fit variables,
$t_i=\pm 1$ (Taylor/Pad\'e) and generically $x^{(\phi)}$ denotes the
physical value of the observable $x$. We perform fits to functions
that all contain the pseudoscalar mass dependencies $v_1=M_\pi^2$,
with $t_1=1$, and $v_2=M_{K_\chi}^2$, with either $t_2=\pm 1$. They
also include the finite-volume term $v_3=M_\pi^{1/2}L^{-3/2}e^{-M_\pi
  L}$, with $t_3=1$. In addition, they contain either of the two
next-to-leading-order terms in $M_\pi^2$, namely $v_4=M_\pi^3$ with
$t_4=1$, or $v_5=M_\pi^4$ with $t_5=1$, corresponding to either a
chiral or generic Taylor expansion in $M_\pi^2$
\cite{Durr:2008zz}. They further include either no discretization term
or the formally leading discretization terms $v_6=\alpha_s a (M_\pi^2 - M_\pi^{(\phi) 2})$ with
$t_6=1$ and $v_7=\alpha_s a (M_{K_\chi}^2 - M_{K_\chi}^{(\phi) 2})$ with $t_7=1$; or the formally
subleading $v_8=a^2 (M_\pi^2 - M_\pi^{(\phi) 2})$, $t_8=1$, and $v_9=a^2 (M_{K_\chi}^2 - M_{K_\chi}^{(\phi) 2})$,
$t_9=1$ \cite{Durr:2008zz}. Note that we set the scale with $M_N$, 
so that discretization terms proportional to just $\alpha_s a$ or $a^2$
would be redundant.

\subsection{Systematic error variations}

\begin{figure}
  \includegraphics[width=0.8\columnwidth]{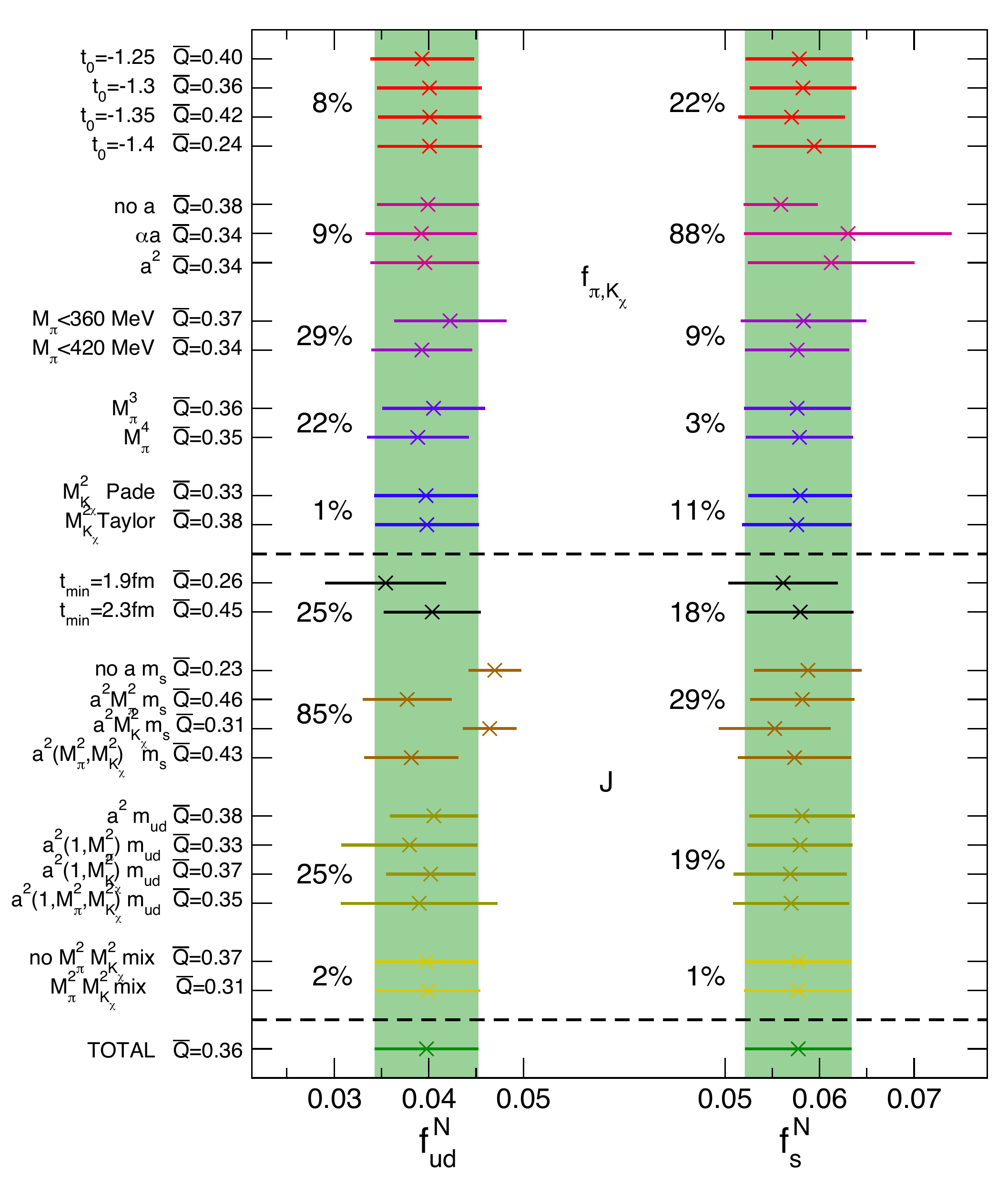}
  \caption{Variation of $f_{ud}^N$ and $f_s^N$ from different
    restrictions on the full analysis procedure. The top part of the
    panel corresponds to variations in the determination of the meson
    mass dependence of the nucleon mass. From top to bottom these are:
  restriction to a single plateau range, to a single scaling
  behavior, to a single pion cut, to either Taylor or chiral expansion
in $M_\pi^2$ and to Taylor or Pad\'e expansion in $M_{K_\chi}^2$. The
middle part displays the effect of variations in the computation of the
elements of the mixing matrix $J$. From top to bottom these are:
restriction to a single plateau range, to a single continuum fit form
for $m_s$, to a single continuum fit form for $m_{ud}$ and to the
inclusion/exclusion of crossterms in the $m_{ud}$, $m_s$ and scale
setting fits. The last row shows the final result
including all fits. The average fit quality of the analyses $\bar{Q}$
is given in each case. The percentage of the total systematic error
due to  each variation alone is also displayed (percentages add up to
$1$ in quadrature).}
  \label{fig:syserr}
\end{figure}

The total number of distinct analyses is 4(plateau
ranges)$\times$2($M_\pi$ cut)$\times$2(chiral/Taylor expansion in
$M_\pi^2$)$\times$2(Taylor/Pad\'e in
$M_{K_\chi}^2$)$\times$3(discretization)$=96$. The way in which we
combine these analyses to give a final central value and systematic
error is described in \sec{sec:uds_res}. In \fig{fig:syserr} we
present the variation of our final observables, $f_{ud}^N$ and
$f_s^N$, that result from applying these different fit
procedures. Results from all different fit procedures are in good
agreement. The leading sources of systematic error come from the
continuum limit and the pion mass cuts.

\subsection{Crosschecks}
From the nucleon fits we can extract the lattice spacings, which are
reported in \tab{tab:latspac}. A comparison to
\cite{Borsanyi:2014jba}, where $M_\Omega$ was used to set the scale on
a set of ensembles which has a large overlap with the current one,
reveals perfect agreement.

\begin{table}[h]
\begin{center}
  \begin{tabular}{cc}
\hline
\hline
    $\beta$ & $a [\mathrm{fm}]$\\
\hline
    3.2 & 0.1022(6)(8) \\
    3.3 & 0.0888(5)(5) \\
    3.4 & 0.0769(5)(4) \\
    3.5 & 0.0637(3)(3) \\
\hline
\hline
  \end{tabular}
\end{center}
  \caption{Lattice spacings obtained from the nucleon fit. Here and
    throughout the paper the first error is statistical and the
    second, the total systematic error, unless stated otherwise.
  \label{tab:latspac}}
\end{table}

To check the validity of our finite-volume ansatz, we verified that
the fit coefficient $c_3=35(13)(5)\mathrm{GeV}^{-2}$ is compatible
with the numerical predictions of \cite{Colangelo:2010ba}. To further
investigate possible finite-volume effects beyond the leading order,
we performed two complete auxiliary analyses with explicit pion
finite-volume effects according to \cite{Colangelo:2005gd}, one with
fixed and one with fitted prefactors. The influence on central values
and errors was found to be insignificant and, in the case of the
fitted prefactor, the additional term was found to be compatible with
zero.

\begin{figure}[!htb]
  \includegraphics[width=\columnwidth]{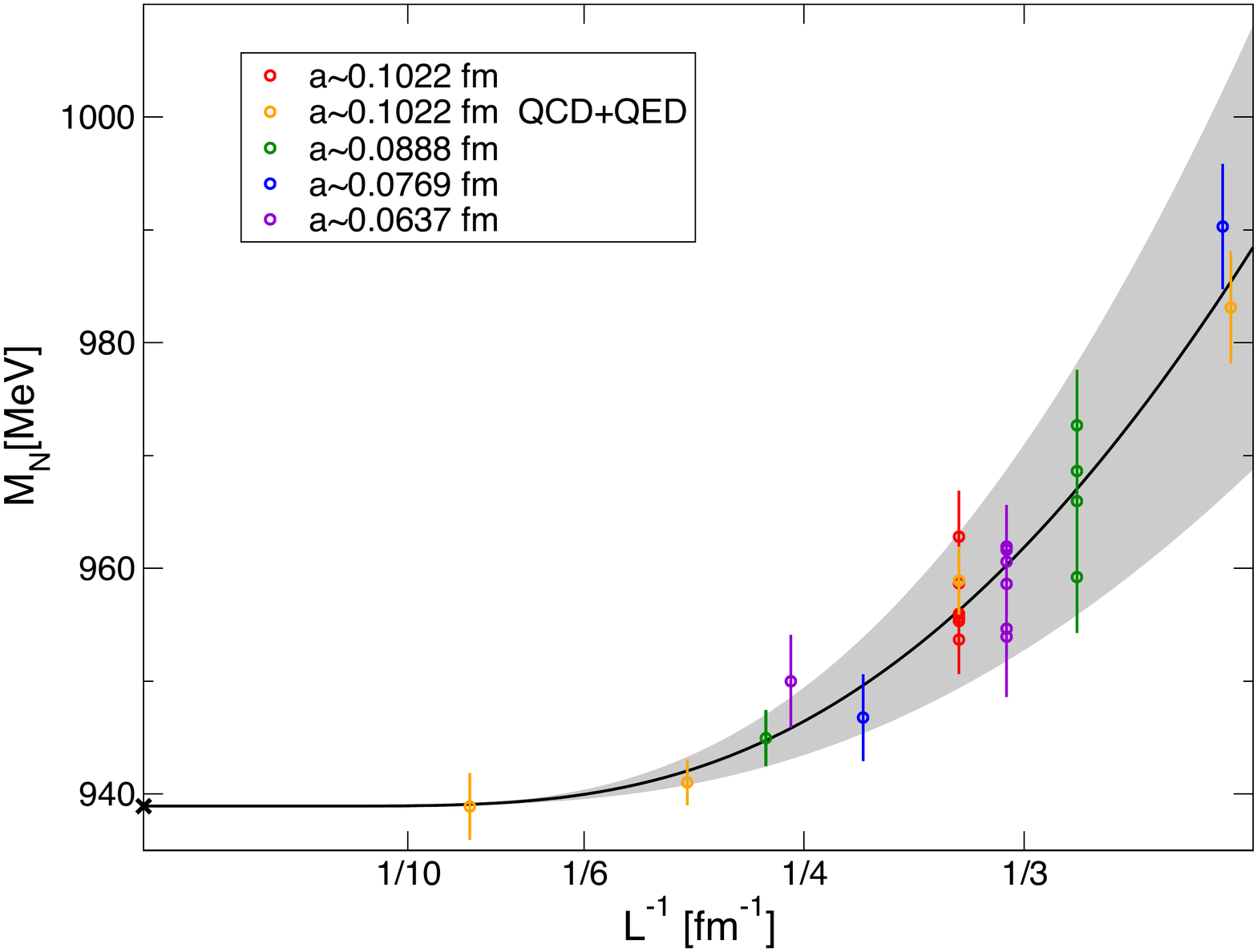}
  \caption{Lattice-spatial-extent dependence of the nucleon mass in
    one sample fit. The fit has been used to shift the simulation
    results to the physical point in all of the other variables on
    which these results depend.}
  \label{fig:nucli}
\end{figure}

\subsection{Results}
\label{sec:uds_res}
The nucleon fits have fit qualities in the range $Q=0.07-0.38$, with an
average fit quality $\bar Q=0.23$. In Figs.~1 and \ref{fig:nucli} we display,
for one sample fit, the dependence of the nucleon mass on $M_\pi^2$,
$M_{K_\chi}^2$ and the inverse of the spatial lattice extent $L$. Defining
pion and reduced kaon $\sigma$-terms as,
\begin{equation}
\sigma_{\pi N}=M_\pi^2\left.\frac{\partial M_N}{\partial M_\pi^2}\right|_{M_{K_\chi}}
\qquad\mbox{and}\qquad
\sigma_{K_\chi N}=M_{K_\chi}^2\left.\frac{\partial M_N}{\partial M_{K_\chi}^2}\right|_{M_\pi}\ ,
\end{equation}
we obtain $\sigma_{\pi N}=42.0(1.3)(1.4)\,\mev$ and
$\sigma_{K_\chi N}=50.9(3.3)(2.8)\,\mev$. We also define the
logarithmic derivatives,
\begin{equation}
\label{eq:logder}
F_\pi^N=\frac{\sigma_{\pi N}}{M_N}=\left.\frac{\partial\ln M_N}{\partial\ln M_\pi^2}\right|_{M_{K_\chi}}
\qquad\mbox{and}\qquad
F_{K_\chi}^N=\frac{\sigma_{K_\chi N}}{M_N}=\left.\frac{\partial\ln M_N}{\partial\ln M_{K_\chi}^2}\right|_{M_\pi}\ ,
\end{equation}
for which our final results are $F_{\pi}^N=0.0447(14)(15)$ and
$F_{K_\chi}^N=0.0542(36)(29)$.  From these fits we also obtain an
estimate of the nucleon mass at vanishing $M_\pi$ keeping $M_{K_\chi}$
fixed, $\left.M_N\right|_{M_\pi=0}=896(1)(2) \,\mev$, and, conversely,
an estimate of the nucleon mass at vanishing $M_{K_\chi}$ keeping $M_\pi$ fixed,
$\left.M_N\right|_{M_{K_\chi}=0}=889(3)(3) \,\mev$. These results, as well as
our value for the nucleon mass in the $SU(3)$ chiral limit
$\left.M_N\right|_{M_\pi=M_{K_\chi}=0}=848(3)(3) \,\mev$, should
be interpreted with great care, since it is not clear that our
interpolation and extrapolation functions are valid down to vanishing quark
masses.

\section{Relation between quark and meson masses}

To leading order in chiral perturbation theory, $M_\pi^2$ and
$M_{K_\chi}^2$ are proportional to $m_{ud}$ and $m_s$. The logarithmic
derivatives $F_{\pi}^N$ and $F_{K_\chi}^N$ defined in
(\ref{eq:logder}) are therefore equal to the nucleon quark content,
\begin{equation}
f_{ud}=\left.\frac{\partial\ln M_N}{\partial\ln m_{ud}}\right|_{m_{s}}
\qquad\mbox{and}\qquad
f_{s}=\left.\frac{\partial\ln M_N}{\partial\ln m_{s}}\right|_{m_{ud}}
\end{equation}
in leading order. To go beyond leading order, we need to determine the
relation between pseudoscalar meson masses and quark masses around the
physical point. For this purpose we use the $N_f=2+1+1$ staggered
ensembles presented in \sec{sec:stagc}, which bracket the physical
point.

\subsection{The Jacobian matrix}

Transforming the $F_P^N$, $P\in\{\pi,K_\chi\}$ into the $f_q$,
$q\in\{ud,s\}$ is achieved via the Jacobian matrix, $J$, with elements
\begin{equation}
\label{eq:jac}
J_{P,q}=\frac{\partial\ln M_P^2}{\partial\ln m_q}\ ,
\end{equation}
evaluated at the physical point so that
\begin{equation}
f_{q}=\frac{\partial\ln M_N}{\partial\ln m_{q}}
=
\sum_P \frac{\partial\ln M_N}{\partial\ln M_P^2}
J_{P,q}
=
\sum_P F_P^N
J_{P,q}
\ .
\end{equation}
While the $F_P^N$ are defined at the physical point, including the
charm quark mass, the $N_f=2+1+1$ ensembles, on which we calculate the
Jacobian, are at a variety of different charm quark masses. Thus, when
discussing the Jacobian, we are careful about recording $m_c$
dependence. Since the computation of this Jacobian is best done with
staggered fermions, whose masses are multiplicatively renormalized, we
parametrize this dependence by the ratio $r_{cs}=m_c/m_s$, where
renormalization factors cancel in mass-independent schemes. Then we
fit the light and strange quark mass as functions of $M_\pi^2$,
$M_{K_\chi}^2$ and $r_{cs}$:
\begin{equation}
  \label{eq:mudmsfn}
m_{ud}(M_\pi^2, M_{K_\chi}^2,r_{cs})
\qquad\mbox{and}\qquad
m_{s}(M_\pi^2, M_{K_\chi}^2,r_{cs}) \ .
\end{equation}
From these fit functions, we determine the following logarithmic derivatives at
the physical point:
\begin{equation}
\label{eq:N}
N_{q,\pi}=\left.\frac{\partial\ln m_{q}}{\partial\ln M_\pi^2}\right|_{M_{K_\chi}^2,r_{cs}}\ ,
\quad
N_{q,K_\chi}=\left.\frac{\partial\ln m_{q}}{\partial\ln M_{K_\chi}^2}\right|_{M_\pi^2,r_{cs}}\ ,
\quad
N_{q,r}=\left.\frac{\partial\ln m_{q}}{\partial\ln r_{cs}}\right|_{M_\pi,M_{K_\chi}^2}\ .
\end{equation}
In principle we need to compute the Jacobian between the two sets of
parameters $(M_\pi^2, M_{K_\chi}^2,r_{cs})$ and
$(m_{ud},m_s,m_c)$. However, the elements of the Jacobian matrix that we are ultimately interested in
are those at fixed $m_c$, i.e.
\begin{equation}
J_{P,ud}=\left.\frac{\partial\ln M_P^2}{\partial\ln
    m_{ud}}\right|_{m_s,m_c}
\qquad\mbox{and}\qquad
J_{P,s}=\left.\frac{\partial\ln M_P^2}{\partial\ln
    m_{s}}\right|_{m_{ud},m_c}\ .
\end{equation}
They are obtained from (\ref{eq:N}) as
\begin{equation}
\begin{split}
J_{\pi,ud}&=\frac{N_{s,K_\chi}}{N_{ud,\pi} N_{s,K_\chi}-N_{ud,K_\chi}N_{s,\pi}}\ ,
\\
J_{\pi,s}&=-\frac{N_{ud,K_\chi}(1-N_{s,r})-N_{s,K_\chi}N_{ud,r}}{N_{ud,\pi} N_{s,K_\chi}-N_{ud,K_\chi}N_{s,\pi}}\ ,
\\
J_{K_\chi,ud}&=-\frac{N_{s,\pi}}{N_{ud,\pi} N_{s,K_\chi}-N_{ud,K_\chi}N_{s,\pi}}\ ,
\\
J_{K_\chi,s}&=\frac{N_{ud,\pi}(1-N_{s,r})-N_{s,\pi}N_{ud,r}}{N_{ud,\pi} N_{s,K_\chi}-N_{ud,K_\chi}N_{s,\pi}}\ .
\end{split}
\end{equation}

\subsection{Renormalization}

The matrix $J$ is a scheme and scale independent quantity. Each
element has the form of $m_q/M_P^2 \times \partial M_P^2 / \partial
m_q$, where $m_q$ is a quark mass and $M_P$, either $M_{\pi^+}$ or
$M_{K_\chi}$. Any multiplicative renormalization factor cancels in
these ratios provided that it is independent of the quark
masses. However, to allow for a global fit of $m_{ud}$ and $m_s$ to
the functions of (\ref{eq:mudmsfn}), involving ensembles with different
lattice spacings, a specific scheme has to be introduced. We
chose the simplest possible scheme: fixing the value of the renormalized
strange quark mass at the physical point to a constant independent of quark
masses and lattice spacing. The exact procedure is detailed in
\sec{sect:ff}

\subsection{Fit functions}
\label{sect:ff}

In order to set the scale, we interpolate the pion decay constant,
$f_\pi$, to the physical mass point defined by the physical values,
$M_\pi^{(\phi)2}$ and $M_{K_\chi}^{(\phi)2}$, of the squared pion and
reduced kaon masses, and by $r_{cs}^{(\phi)}$, which we set to
$11.85$. Since all of our results are dimensionless ratios, scale
setting is not critical and especially a variation in
$r_{cs}^{(\phi)}$ has negligible effect on the result. Varying the
interpolation function has no effect on our results as is evident from
\fig{fig:syserr}.

To determine $J$ we study the behavior of the functions
$m_{ud}(M_\pi^2, M_{K_\chi}^2,r_{cs})$ and $m_{s}(M_\pi^2,$ $M_{K_\chi}^2$, $r_{cs})$.
We expand these around the physical mass point, again defined by
$M_\pi^{(\phi)2}$, $M_{K_\chi}^{(\phi)2}$ and $r_{cs}^{(\phi)}$. The fit
functions we employ have the form
\begin{equation}
  m_{ud}^R(M_\pi^2, M_{K_\chi}^2,r_{cs}) = 
    c_{00}^{ud}
    + c_{10}^{ud}\Delta_\pi
    + c_{01}^{ud}\Delta_{K_\chi}
    + c_{20}^{ud}\Delta_\pi^2
    + c_{02}^{ud}\Delta_{K_\chi}^2
    + c_{11}^{ud}\Delta_\pi\Delta_{K_\chi}
    + c_{c1}^{ud}\Delta_{r_{cs}}
\end{equation} 
and
\begin{equation}
  m_{s}^R(M_\pi^2, M_{K_\chi}^2,r_{cs}) = 
    c_{00}^{s}
    + c_{10}^{s}\Delta_\pi
    + c_{01}^{s}\Delta_{K_\chi}
    + c_{20}^{s}\Delta_\pi^2
    + c_{02}^{s}\Delta_{K_\chi}^2
    + c_{11}^{s}\Delta_\pi\Delta_{K_\chi}
    + c_{c1}^{s}\Delta_{r_{cs}}\ ,
\end{equation} 
where $\Delta_\pi = M_\pi^{2} - M_\pi^{(\phi)2}$, $\Delta_{K_\chi}
= M_{K_\chi}^{2} - M_{K_\chi}^{(\phi)2}$, $\Delta_{r_{cs}}=r_{cs}-r_{cs}^{(\phi)}$ and $m_q^R$ are renormalized
quark masses. They are related to the bare quark masses $m_q$, which
are parameters of the action, by a multiplicative renormalization
factor, $Z$, so that $m_{ud}^R = Z m_{ud}(M_\pi^2, M_{K_\chi}^2,r_{cs})$ and
$m_{s}^R = Z m_s(M_\pi^2, M_{K_\chi}^2,r_{cs})$. We choose a scheme in which
the renormalization factors do not depend on the quark masses, and
consequently the squared pion and reduced kaon masses, but only on the
gauge coupling $\beta$. To avoid an explicit determination of the
renormalization factors, we divide the above expansions by the value
of the renormalized strange quark mass at the physical mass point for
each $\beta$, yielding:
\begin{equation}
  \label{eq:mudovmsff}
  \frac{m_{ud}(M_\pi^2, M_{K_\chi}^2,r_{cs})}{m_{s}(M_\pi^{(\phi)2}, M_{K_\chi}^{(\phi)2},r_{cs}^{(\phi)})} = 
    r
    + d_{10}^{ud}\Delta_\pi
    + d_{01}^{ud}\Delta_{K_\chi}
    + d_{20}^{ud}\Delta_\pi^2
    + d_{02}^{ud}\Delta_{K_\chi}^2
    + d_{11}^{ud}\Delta_\pi\Delta_{K_\chi}
    + d_{c1}^{ud}\Delta_{r_{cs}}
\end{equation}  
and
\begin{equation}
  \label{eq:msovmsff}
  \frac{m_{s}(M_\pi^2, M_{K_\chi}^2,r_{cs})}{m_{s}(M_\pi^{(\phi)2}, M_{K_\chi}^{(\phi)2},r_{cs}^{(\phi)})} = 
    1
    + d_{10}^{s}\Delta_\pi
    + d_{01}^{s}\Delta_{K_\chi}
    + d_{20}^{s}\Delta_\pi^2
    + d_{02}^{s}\Delta_{K_\chi}^2
    + d_{11}^{s}\Delta_\pi\Delta_{K_\chi}
    + d_{c1}^{s}\Delta_{r_{cs}}\ ,
\end{equation}
where $r$ is the ratio of the strange and to light quark mass at the
physical point and $d_{ij}^q = c_{ij}^q / m_{s}(M_\pi^{(\phi)2},
M_{K_\chi}^{(\phi)2},r_{cs}^{(\phi)})$, for $q=ud,s$ and for all values of $ij$
appearing in \eqs{eq:mudovmsff}{eq:msovmsff}. Note that all
renormalization factors cancel out. The value $m_{s}(M_\pi^{(\phi)2},
M_{K_\chi}^{(\phi)2},r_{cs}^{(\phi)})$ is different for each gauge coupling
$\beta$. This is a manifestation of the need for renormalization. In
practice, we introduce fit parameters $m_s^{(\phi)}[\beta]$, where
$\beta$ in rectangular brackets indicates that one independent
parameter per gauge coupling is considered.

The above expansions are subject to discretization artefacts. We consider
these to contribute to the ratio $r$ and to the linear terms in
$\Delta_\pi$ and $\Delta s$. Discretization artefacts on the quadratic terms
are subleading. The final fit functions are 
\begin{eqnarray}
\label{eq:mud}
  \frac{m_{ud}(M_\pi^2, M_{K_\chi}^2,r_{cs})}{m_s^{(\phi)}[\beta]} & = & 
    r_0 + a^2 r_1
    + (d_{10}^{ud} + e_{10}^{ud} a^2)\Delta_\pi 
    + (d_{01}^{ud} + e_{01}^{ud} a^2)\Delta_{K_\chi} + \nonumber \\ & &
      d_{20}^{ud}\Delta_\pi^2
    + d_{02}^{ud}\Delta_{K_\chi}^2
    + d_{11}^{ud}\Delta_\pi\Delta_{K_\chi}
    + d_{c1}^{ud}\Delta_{r_{cs}}
\end{eqnarray}  
and
\begin{eqnarray}
  \frac{m_{s}(M_\pi^2, M_{K_\chi}^2,r_{cs})}{m_s^{(\phi)}[\beta]} & = & 
    1
    + (d_{10}^{s} + e_{10}^{s} a^2)\Delta_\pi
    + (d_{10}^{s} + e_{10}^{s} a^2)\Delta_{K_\chi} + \nonumber \\ & &
      d_{20}^{s}\Delta_\pi^2
    + d_{02}^{s}\Delta_{K_\chi}^2
    + d_{11}^{s}\Delta_\pi\Delta_{K_\chi}
    + d_{c1}^{s}\Delta_{r_{cs}}.
\end{eqnarray}
We perform various fits, all of which contain the fit parameters
$r_0$, $r_1$, $d_{10}^q, d_{01}^q$ and $ d_{c1}^q$. The discretization terms
$e_{10}^q$, $e_{01}^q$ and the higher order interpolation terms
$d_{11}^{q}$ are optionally present and we perform fits with all
possible variations, with each term either present or absent. The
additional, higher-order terms, $d_{20}^{q}$ and $d_{02}^{q}$, turn
out to be irrelevant and can be omitted entirely.

Finite-volume effects on all relevant quantities are safely below 1
permil on all of our ensembles \cite{Colangelo:2005gd} and cannot be
detected with the statistical accuracy of our data.

\begin{figure}
  \includegraphics[width=\columnwidth]{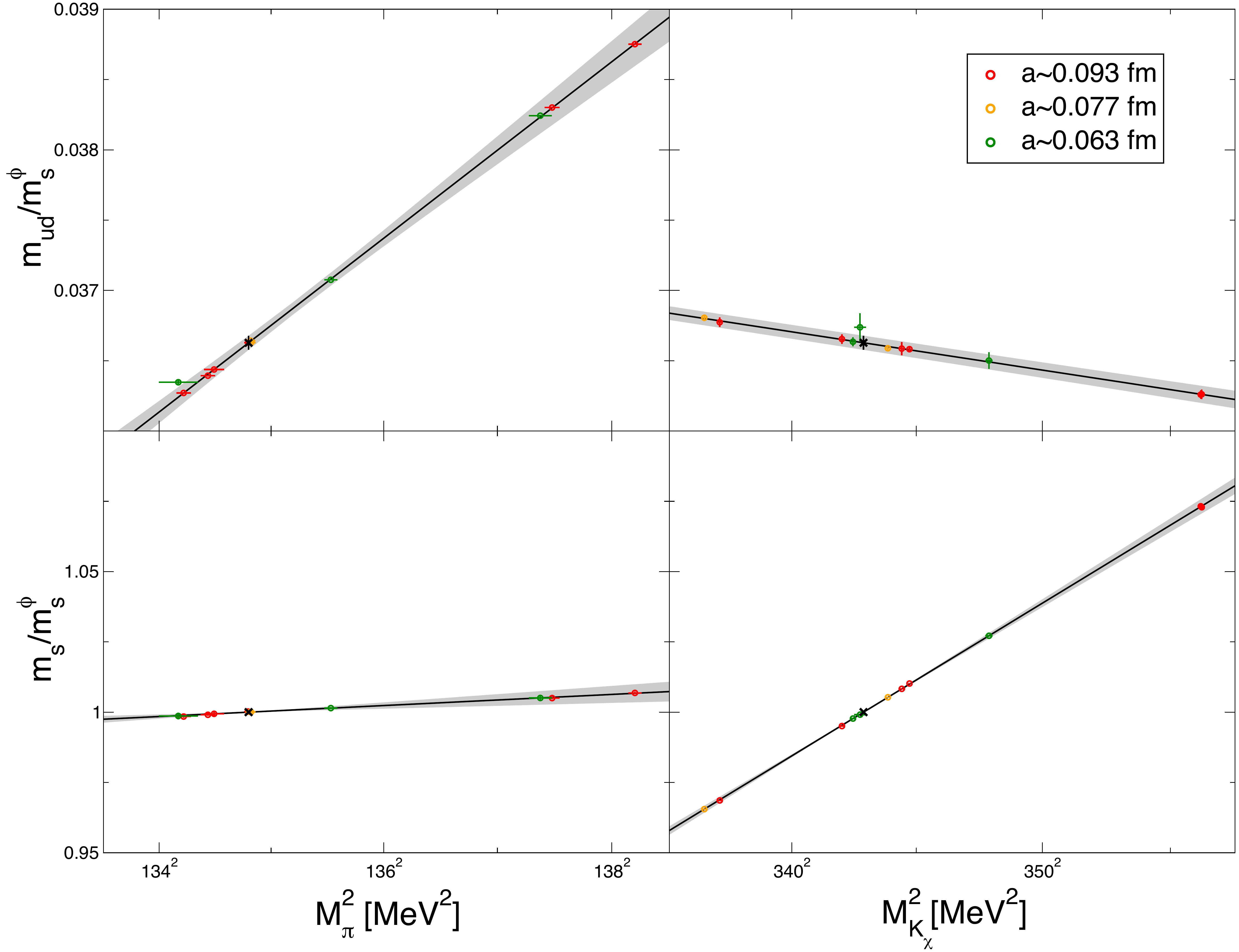}
  \caption{\label{fig:mixm}Dependence of the quark masses, $m_{ud}$
    and $m_s$, on the squared pseudoscalar-meson masses, $M_\pi^2$ and
    $M_{K_\chi}^2$, for one of our analyses. Data points are corrected
    using the fit to tune all, except the plotted variables, to their
    physical value.  }
\end{figure}

The $M_\pi^2$ and $M_{K_\chi}^2$ dependence of the quark masses for a
single representative analysis are displayed in \fig{fig:mixm}.

\subsection{Systematic error variations}

The total number of distinct analyses for the Jacobi matrix is
2(plateau ranges) $\times$ 4(discretization terms on $m_s$) $\times$
4(discretization terms on $m_{ud}$) $\times$ 2(higher order
interpolation terms) = 64. The way in which we
combine these analyses to give a final central value and systematic
error is described in \sec{sec:uds_res}. In \fig{fig:syserr} we present the
variation of our final observables, $f_{ud}^N$ and $f_s^N$, resulting from
these different fit procedures. Results from all
fit procedures are in good agreement, with the leading contribution
towards the systematic error coming from the variation of the continuum
extrapolation terms in $m_s$.

\subsection{Strange to light quark mass ratio}

\begin{figure}
  \includegraphics[width=0.8\columnwidth]{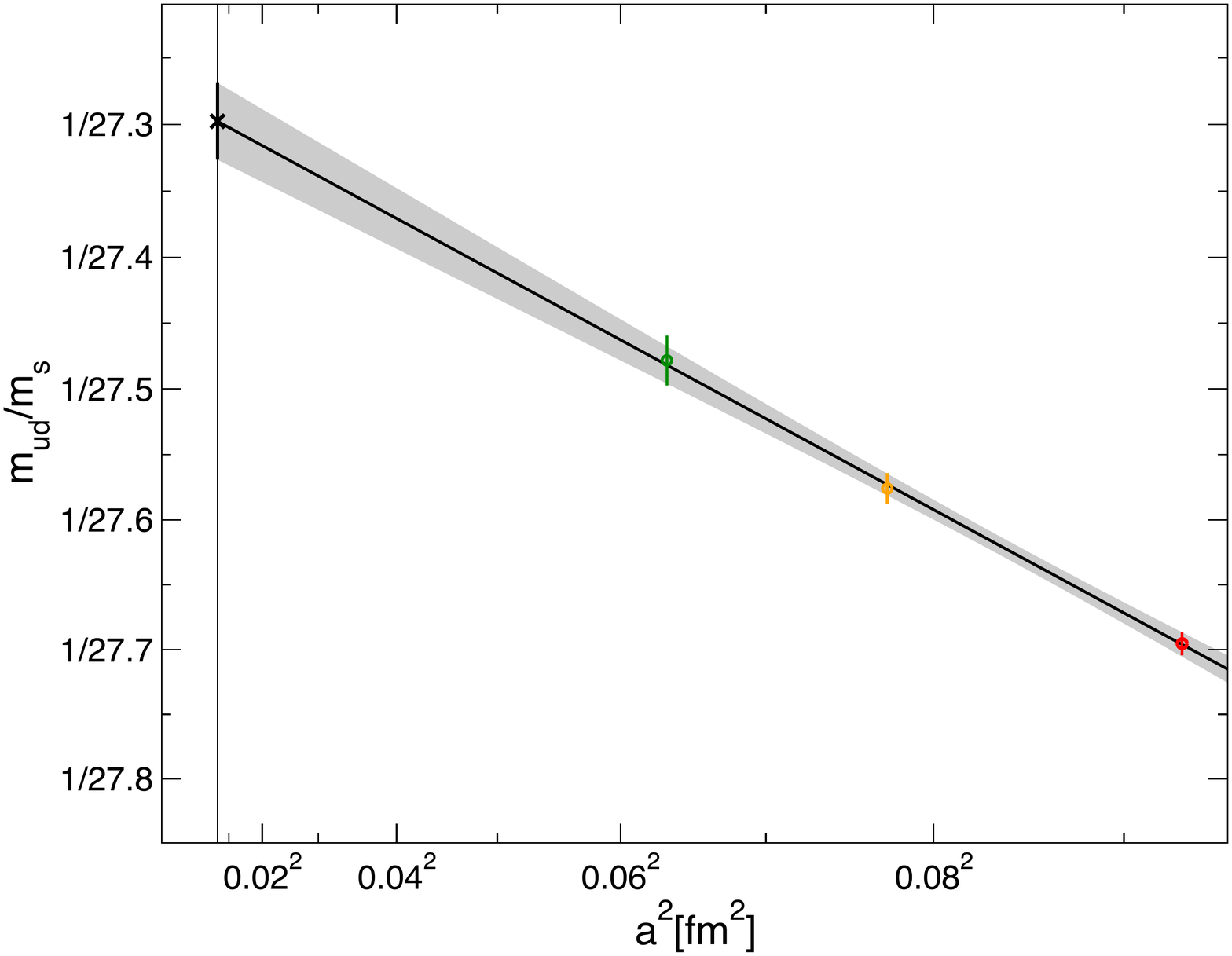}
  \caption{\label{fig:msmud}Example of a continuum extrapolation of
    the quark mass ratio $m_s/m_{ud}$.  }
\end{figure}

One of the fit parameters in (\ref{eq:mud}), $r_0$, is the ratio of
light to strange quark masses at the physical point. This is a
phenomenologically interesting parameter that we can determine. It
provides an additional crosscheck of our fit procedure. For its
inverse, the strange to light quark mass ratio, we obtain
\begin{equation}
\frac{m_s}{m_{ud}}=27.29(33)(8)
\end{equation}
which is in good agreement with the current PDG world average
$27.3(7)$ from \cite{Tanabashi:2018oca} and with the most precise
lattice values $27.35(5)({+10\atop-7})$ from \cite{Bazavov:2014wgs}
and $27.53(20)(8)$ from \cite{Durr:2010vn}. The continuum
extrapolation of this quantity for a single representative analysis is
shown in \fig{fig:msmud}.

\subsection{Results}

\begin{figure}
  \includegraphics[width=\columnwidth]{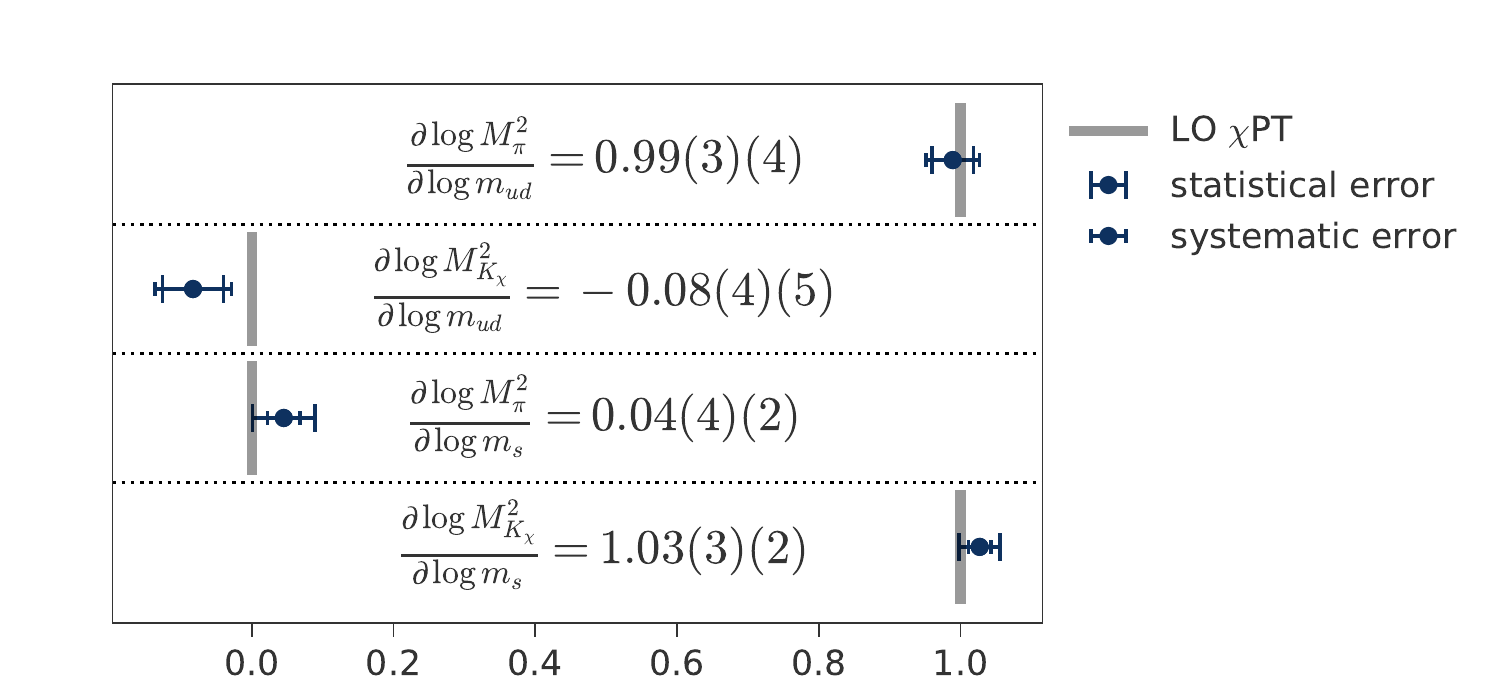}
  \caption{\label{fig:jac} Final results for the elements of the
    Jacobian matrix (\ref{eq:jac}). There are only small corrections to
    the leading order $\chi$PT predictions.}
\end{figure}

The quark mass fits have fit qualities in the range $Q=0.05-0.99$,
with an average fit quality of $Q=0.62$.  The final results on the
elements of the mixing matrix are given in \fig{fig:jac}. Evidently,
the mixing matrix provides only a small correction to the leading
order prediction of chiral perturbation theory, where the mixing
matrix is the identity.

\begin{figure}
  \includegraphics[width=0.8\columnwidth]{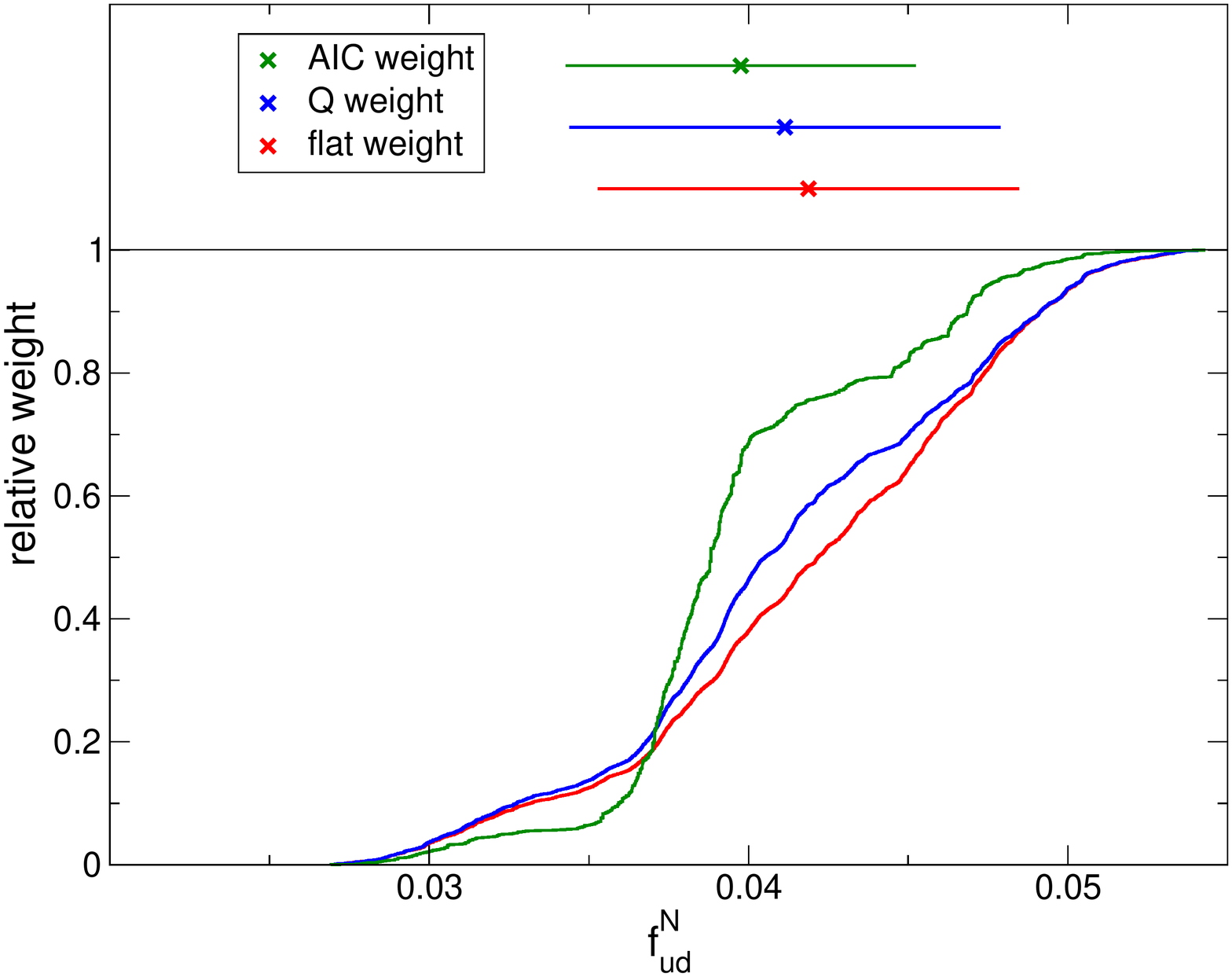}
\caption{\label{fig:fudcdf} The cumulative distribution function of
  $f^N_{ud}$ over all analyses, obtained using three different weight
  functions. In the top part of the figure, the resulting central values
  and total errors, corresponding to each weight function, are
  plotted.  }
\end{figure}

Applying the Jacobian matrix to the vector, $(F_{\pi^+}^N,
F_{K_\chi}^N)^T$, we obtain the final results listed in Table~1. We
use a weighted average and standard deviation of the results from the
individual analyses to determine central value and systematic errors
\cite{Durr:2008zz,Borsanyi:2014jba}. For our main result, we use the
Akaike information criterion (AIC) \cite{Akaike:1974} to determine the
relative weight of the analyses. In order to check that the choice of
weight does not significantly alter the result, we have plotted the
cumulative distribution function of $f^N_{ud}$ and $f^N_s$ in
\figs{fig:fudcdf}{fig:fmscdf}, with a flat, unit weight, with a weight
equal to the quality of fit, $Q$, and with the AIC weight.  The choice
of weight does not substantially influence the final result.

\begin{figure}
  \includegraphics[width=0.8\columnwidth]{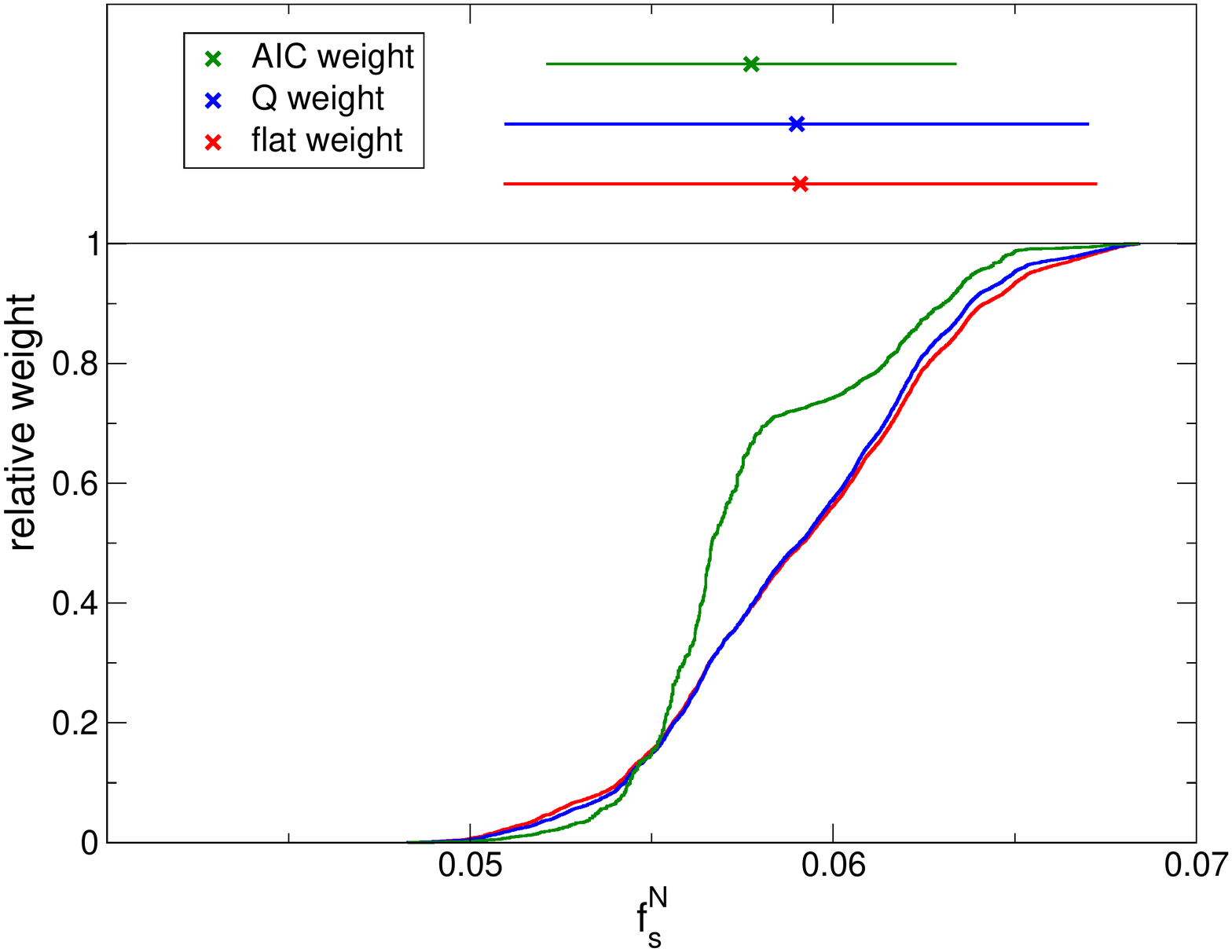}
\caption{\label{fig:fmscdf}
The cumulative distribution function of
  $f^N_{s}$ over all analyses, obtained using three different weight
  functions. In the top part of the figure, the resulting central values
  and total errors, corresponding to each weight function, are
  plotted. 
}
\end{figure}

Our results are largely compatible with other recent lattice
determinations \cite{Durr:2015dna,Bali:2016lvx,Alexandrou:2017xwd,Alexandrou:2017qyt,Ling:2017jyz,Alexandrou:2019brg} as well as with
the seminal calculations by Gasser, Leutwyler and Sainio
\cite{Gasser:1990ce}, while recent phenomenological determinations
that obtain $f^N_{ud}$ from $\pi N$ scattering data
\cite{Pavan:2001wz,Alarcon:2011zs,Chen:2012nx,Hoferichter:2015dsa,Hoferichter:2015hva} give somewhat
higher values.

\section{Heavy quark effective theory}

Scalar quark contents, the nucleon mass and the QCD trace anomaly are
related by a sum rule \cite{Shifman:1978zn}. This sum rule allows one
to compute the heavy-quark contents from the light-quark ones, in the
heavy-quark limit. This is achieved by considering a succession of
effective field theories of QCD in which the heavy top, bottom and
charm quarks are integrated out in turn. Thus, consider QCD with $N_f$
flavors treated as light and with one heavy quark, $Q$, to be
integrated out. Then, matching the $N_f+1$ and $N_f$ theories at scale
$m_Q$ yields\cite{Shifman:1978zn} yields
\begin{equation}
  \label{eq:hqelo}
  f^N_Q=\frac{2}{3\beta_0}(1-\bar f_{N_f})\left[1+O\left(\left(\frac{\Lambda_{QCD}}{m_Q}\right)^2,
    \alpha_s(m_Q)\right)\right]
\end{equation}
where $\beta_0=11-(2/3)N_f$ is the leading-order coefficient of the QCD
$\beta$-function, in $N_f$-flavor QCD, and
\begin{equation}
\label{eq:lnf}
  \bar f_{N_f}=\sum_{i=1}^{N_f}f^N_{q_i}
\end{equation}
is the sum of the scalar quark contents over the $N_f$ quark flavors,
$q_i$, not integrated out. Recently, higher-order QCD corrections to
(\ref{eq:hqelo}) have been computed to $O(\alpha_s^3)$
\cite{Hill:2014yxa}, yielding
\begin{equation}
  f_Q=\sum_{n=0}^3 (b^{N_f}_n-c^{N_f}_n \bar f_{N_f})\alpha_s^n\left[1
   +O\left(\left(\frac{\Lambda_{QCD}}{m_{Q}}\right)^2,\alpha_s^4(m_{Q})\right)\right])
\label{eq:hqe}
\end{equation}
where $N_f=5$ for $Q=t$, $N_f=4$ for $Q=b$ and $N_f=3$ for $Q=c$. The
numerical values of the $b_n$ and $c_n$ are given in \tab{tab:hscoef} for
$N_f=3,4$~and~$5$.

\begin{table}[h]
\begin{center}
  \begin{tabular}{ccccccc}
\hline
\hline
   $n$   & $b_n^3$ & $c_n^3$& $b_n^4$ & $c_n^4$& $b_n^5$ & $c_n^5$\\
\hline
   0 & 0.0740741  & 0.0740741 & 0.08 & 0.08 & 0.0869565  & 0.0869565  \\
   1 & 0.0229236  & 0.0700806 & 0.0308124 &0.081742 & 0.0412178 & 0.0965761 \\
   2 & 0.0412178 & 0.0534895 & 0.0157223 & 0.0664099 & 0.0246729 & 0.0846867\\
   3 & -0.012595   & -0.0245554 & -0.0218678 & -0.0409409 & -0.0334609 & -0.0578502 \\
\hline
\hline
  \end{tabular}
\end{center}
  \caption{Coefficients in the heavy quark expansion (\ref{eq:hqe}) from \cite{Hill:2014yxa}. 
  \label{tab:hscoef}}
\end{table}

We leverage these results in various ways. First, we can obviously use
our results for the light quark contents to determine various
$\bar f_{N_f}$ and thus the heavy-quark contents. This method is used to
compute $f^N_b$ and $f^N_t$ in \sec{sec:btsig}.

Now, if only the top quark is integrated out, the neglected
corrections in (\ref{eq:hqe}) are neglible. If also the bottom quark
is removed from the dynamics, these corrections are still very small
since $(\Lambda_{QCD}/m_b)^2\sim 0.6\%$ ($\alpha_s(m_b)^4$ is
smaller). The heavy-quark expansion is possibly less-well behaved for
the charm. Indeed, $(\Lambda_{QCD}/m_c)^2\sim 6\%$
($\alpha_s(m_c)^4\lsim 2\%$ is smaller), which is no longer negligible
and is comparable to our lattice uncertainties. Thus, we also compute
$f^N_c$ directly on the lattice, using the Feynman-Hellmann theorem. As shown in \sec{sec:fc}, the result
obtained is compatible with the one given by the heavy-quark
expansion, within the naive estimate of $O((\Lambda_{QCD}/m_c)^2)$
corrections. This indicates that, even for the charm, the heavy-quark
expansion behaves as expected.

The heavy-quark expansion also helps in the direct determination of
the scalar, charm-quark content on the lattice. Since the nucleon mass
depends on the charm-quark mass only very mildly, we vary this
mass by $\pm 25\%$ around its physical value and measure the
corresponding variations of the nucleon mass:
\begin{equation}
\begin{matrix}
\Delta_+&=&M_N(m_c=\frac{5}{4}m_c^\phi)-M_N(m_c=m_c^\phi)\ ,\\
\Delta_-&=&M_N(m_c=m_c^\phi)-M_N(m_c=\frac{3}{4}m_c^\phi)\ .
\label{eq:dpm}
\end{matrix}
\end{equation}
These are then combined to determine the quantity
\begin{equation}
\label{eq:tayc}
\hat f^N_c=2\frac{\Delta_++\Delta_-}{M_N(m_c=m_c^\phi)}
\end{equation}
which, using a second order Taylor series expansion in $m_c$ around
the physical point, can be shown to equal the scalar, charm-quark content
$f^N_c$, up to terms of order $(\delta m_c/m_c)^2=1/16$.

Insight from the heavy-quark expansion provides an alternative
expansion. First we note that $\alpha_s$ changes roughly between $0.5$
and $0.34$ when changing the scale from $3/4m_c^\phi$ to $5/4m_c^\phi$
\cite{Baikov:2016tgj}, implying a relative variation of $f^N_c$ in
this range by $\sim3\%$ according to (\ref{eq:hqe}). To a good
approximation $f^N_c(m_c)$, in this range, is therefore constant, i.e.
\begin{equation}
f^N_c(km_c)=
\left.\frac{\partial}{\partial\ln m}\ln M(m)\right|_{m=km_c}\simeq\text{cst}
\end{equation}
for $k=0.75$ or $k=1.25$. This implies
\begin{equation}
M(m_c)=
M(m_c^\phi)
\left(
\frac{
m_c}{
m_c^\phi}
\right)
^{f^N_c}
\ .\end{equation}
Taylor expanding this expression to second order around $f_c=0$ and
plugging the result into (\ref{eq:dpm}), we find that the quantity,
\begin{equation}
\label{eq:hqc} 
\tilde f^N_c=
\frac{1}{
M(m_c^\phi)
\ln\frac 5 4 \ln\frac 4 3 \ln\frac 5 3
}
\left(
\left(
\ln\frac 4 3
\right)^2
\Delta_+
+
\left(
\ln\frac 5 4
\right)^2
\Delta_-
\right)
\ ,\end{equation}
is equal to the charm-quark content $f^N_c$, up to terms of order
$(f^N_c)^3\sim3\times10^{-4}$, assuming $f^N_c$ to be constant in the
$\pm25\%$ region around the physical charm quark mass. Since the
heavy-quark expansion tells us that this assumption is valid to
$O(3\%)$, we conclude that $\tilde f^N_c=f^N_c+O(3\%)$.  The
difference between $\hat f^N_c$ extracted with the Taylor series
(\ref{eq:tayc}) and $\tilde f^N_c$ extracted with the heavy quark
expansion (\ref{eq:hqc}) provides an estimate of the systematic error
due to replacing the derivative of the nucleon mass, with respect to
$m_c$, by a finite difference.

\section{Lattice computation of the $c$ sigma term}
\label{sec:fc}

The direct computation of the charm-quark content from our lattice
ensembles poses a different set of challenges. On the one hand,
locating the physical point precisely is not critical, as detailed in
the previous section. On the other hand, one needs to vary the
charm-quark mass over a significant range to obtain a signal. The
strategy that we employ takes these two considerations into
account. Instead of performing a combined fit to the dependence of the
nucleon mass on lattice spacing, quark masses and volume and using the
result to compute its derivative with respect to $m_c$, we directly
determing the charm-quark content from finite differences
(\ref{eq:tayc},\ref{eq:hqc}) at each lattice spacing. The central
ensembles, at each lattice spacing, are tuned to the physical mass
point to within less than 4 \% in the light and strange quark masses
leading to tiny corrections covered by the variation of
finite-difference forms (\ref{eq:tayc},\ref{eq:hqc}). Furthermore,
since the ensembles at one lattice spacing share all parameters except
for the charm-quark mass, and $M_\pi L>4$, finite-volume effects are
irrelevant too.

\subsection{Analysis details}

\begin{figure}
  \includegraphics[width=0.8\columnwidth]{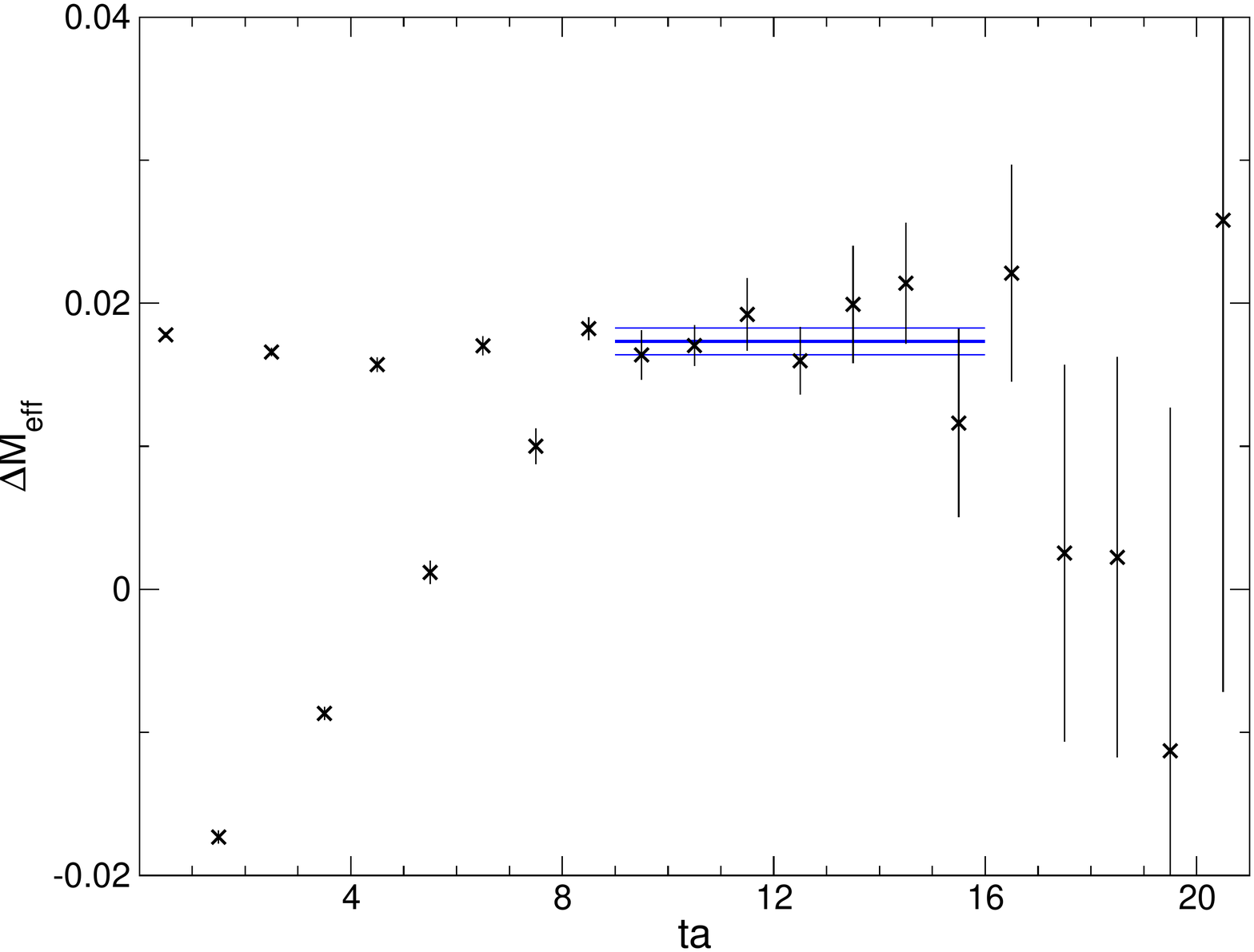}
  \caption{\label{fig:delplat}
Effective mass plateau and extracted mass from the more aggressive
fit with $t_{\mathrm{min}}\sim0.8\,\fm$ at $\beta=3.84$.
}
\end{figure}

We compute the mass differences $\Delta_\pm$ of (\ref{eq:dpm})
directly from the ratio of nucleon correlators from the two relevant
ensembles. Staggered excited states in the two nucleon correlators are
suppressed by using the time-shifted propagator described in
\sec{sect:extract}, for each of the ensembles
separately. \fig{fig:delplat} shows an effective mass plateau for
$\Delta_M=\Delta_++\Delta_-$, for our $\beta=3.84$ ensembles. We
identify the onset of the plateau to be slightly below
$t_{\mathrm{min}}\sim 0.8\,\fm$ and correspondingly consider two
values, $t_{\mathrm{min}}\simeq 0.8\,\fm$ and
$t_{\mathrm{min}}\simeq 0.95\,\fm$, for determining the mass
difference. The variation of the result with respect to
$t_{\mathrm{min}}$ enters into the systematic error.

We determine the lattice spacing either by interpolating the pseudoscalar
decay constant $f_\pi$ as describes in \sec{sect:ff} or by
directly using the nucleon mass from the central, physical ensemble at each
$\beta$. The difference between these two procedures also enters into
the systematic error estimate.

\begin{figure}
  \includegraphics[width=0.8\columnwidth]{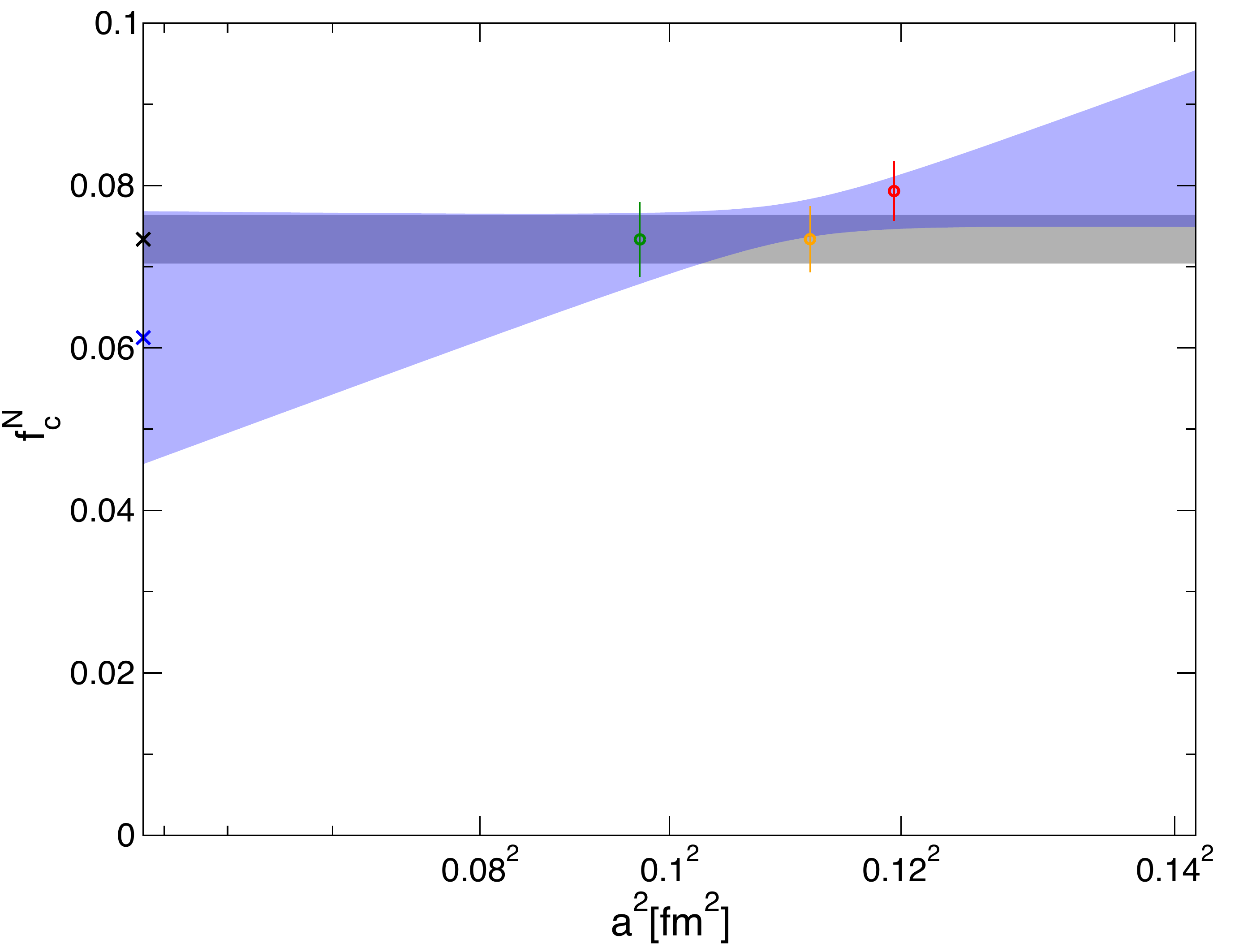}
\caption{\label{fig:charmscal} Continuum extrapolation of the scalar,
  charm-quark content of the nucleon, $f^N_c$. The blue curve and
  point correspond to a linear extrapolation in $a^2$ using all three
  lattice spacings while the grey curve and black point corresponds to
  a constant extrapolation to the results from our two finest
  lattices.  }
\end{figure}

In order to estimate cutoff uncertainties, we perform three different
continuum extrapolations of $f^N_c$. Using values from all three
lattice spacings, we perform either a constant or linear extrapolation
in $a^2$. In addition, we also perform a constant extrapolation using
the two finest lattice spacings only. The result from two of these
extrapolations is plotted in fig.~\ref{fig:charmscal}. The spread
between the results of these methods again enters the systematic error.

\subsection{Systematic error variations}

We perform a total of 2($\hat f^N_c$,$\tilde f^N_c$)$\times$2(fit
range) $\times$2(Scale setting)$\times$3(continuum extrapolation)$=24$
different analysis procedures. The way in which we
combine these analyses to give a final central value and systematic
error is described in \sec{sec:c_res}.

\begin{figure}
  \includegraphics[width=0.8\columnwidth]{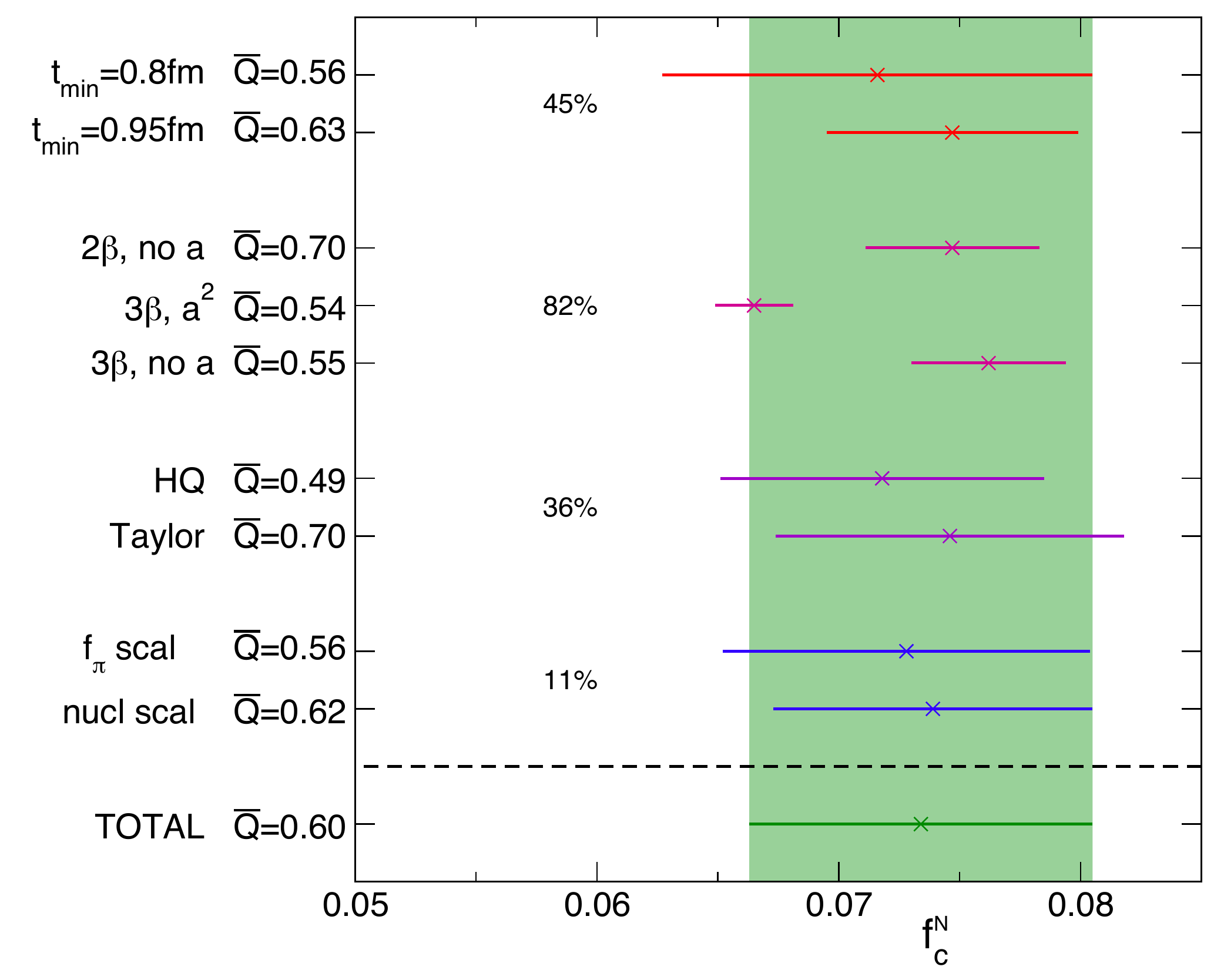}
  \caption{Variation of $f_{c}^N$ from different restrictions to the
    full analysis procedure. From top to bottom these are: restriction
    to a single plateau range, to a single scaling behavior, to a
    single definition for $f^N_c$ and to a single scale-setting
    method. The last row shows the final result including all
    fits. The average fit quality of the analyses $\bar{Q}$ is given
    in each case. The percentage of the total systematic error due to
    each variation alone is also displayed (percentages add up in
    quadrature).}
  \label{fig:syserrc}
\end{figure}

\fig{fig:syserrc} gives a breakup of the systematic error into its
different components. As one can see, all restrictions of the fit
procedure are in agreement with the final result and the main
contribution to the systematic uncertainty originates from varying the
continuum fit.

\subsection{Crosschecks}

As discussed in \sec{sec:fc}, heavy-quark effective theory provides us
with a crosscheck of the charm sigma term up to a precision of
$O((\Lambda_{QCD}/m_c)^2)\sim 6\%$. We enter, into the heavy-quark
expansion (\ref{eq:hqe}), $\alpha_s(m_{c})=0.388(13)$ originating from
$m_c(m_c)=1.275({+25\atop -35})\mathrm{GeV}$ \cite{Tanabashi:2018oca}
and the numerically integrated 5-loop beta function
\cite{Baikov:2016tgj} with $\alpha_s(91.187\mathrm{GeV})=0.1181$
\cite{Tanabashi:2018oca}.  This results in
\begin{equation}
f_c^N|_\text{HQ}=0.08374(34) - 0.1078(14) \bar f_3
\end{equation}
where $\bar f_3$ denotes the sum of quark contents of the lighter quarks
\begin{equation}
\bar f_3=\sum_{q=ud,s}f_{q}^N
\ .
\end{equation}
Our light quark results give
\begin{equation}
\bar f_3=0.0975(56)(52)\ ,
\end{equation}
yielding
\begin{equation}
f_c^N|_\text{HQ}=0.07323(61)(65)
\end{equation}
This result is in excellent agreement with the one from the direct
lattice determination.

\subsection{Results}
\label{sec:c_res}

The continuum extrapolations have fit qualities in the range
$Q=0.25-1.00$, with an average fit quality $\bar Q=0.60$.

\begin{figure}
  \includegraphics[width=0.8\columnwidth]{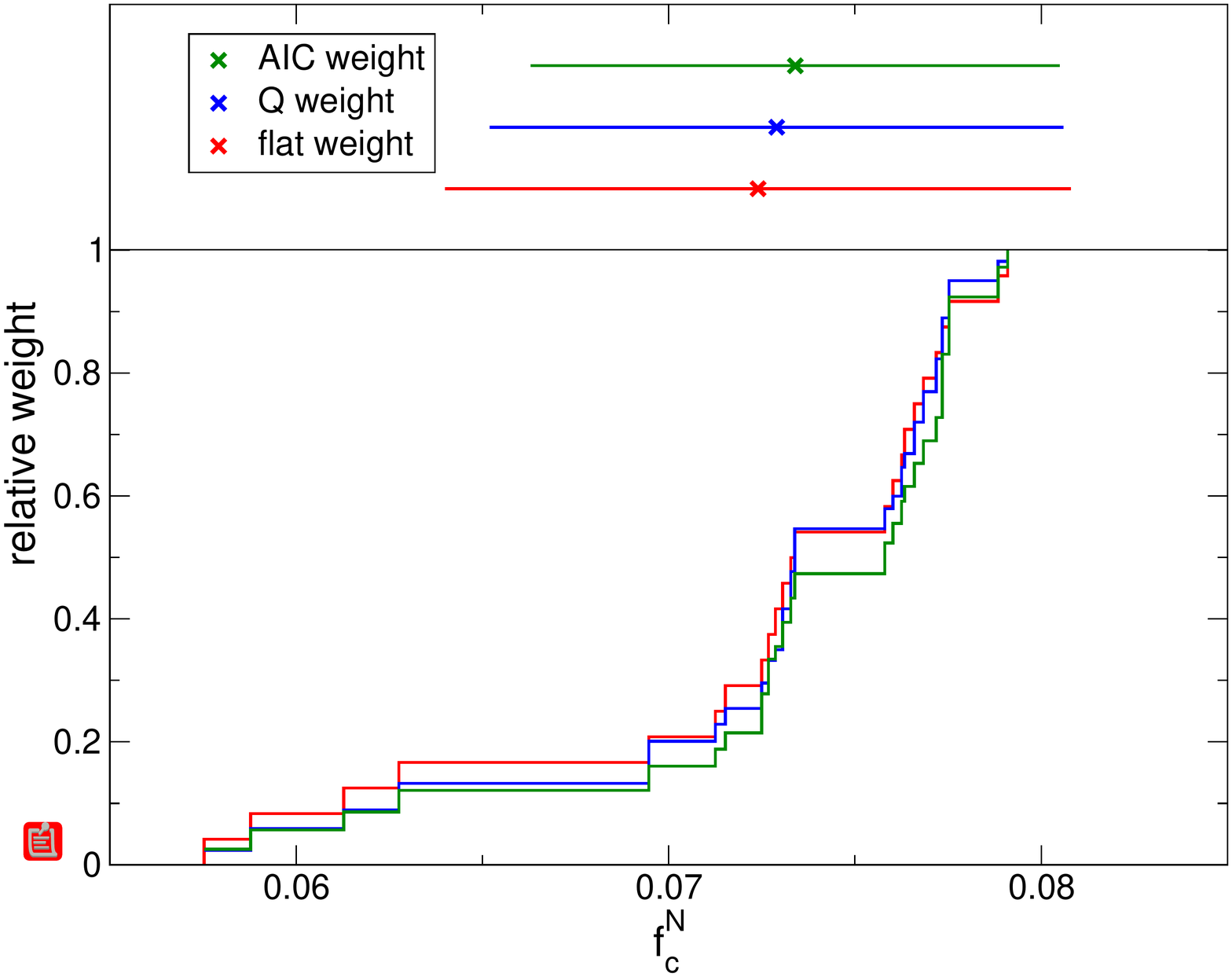}
\caption{\label{fig:fmccdf}
The cumulative distribution function of
  $f^N_{c}$ over all analyses, obtained using three different weight
  functions. In the top part of the figure, the resulting central values
  and total errors, corresponding to each weight function, are
  plotted. 
}
\end{figure}

As in the case of the light and strange quark contents, we plot the
cumulative distribution function of $f^N_c$ from all analyses with
different weight functions (see \fig{fig:fmccdf}). The choice of
the weighting function does not significantly affect our result and we
take the AIC weight for producing our final result.

The lattice calculation of $f^N_c$ \cite{Alexandrou:2017xwd,
  Alexandrou:2017qyt,Alexandrou:2019brg} is compatible with our
number. So is the determination of \cite{Freeman:2012ry}, in which
systematic errors are not estimated.

\section{Heavy-quark expansion for the $b$, and $t$ $\sigma$-terms}
\label{sec:btsig}

Having checked the validity of the heavy-quark expansion
(\ref{eq:hqe}) in the case of the charm, we use it to compute $f^N_b$
and $f^N_t$. We use as inputs
$\alpha_s(m_{b/t})=0.225(3)/0.109(1)$ originating from
$m_b(m_b)=4.18({+4\atop -3})\mathrm{GeV}$ and
$m_t(m_t)=160.0({+4.8\atop -4.3})\mathrm{GeV}$
\cite{Tanabashi:2018oca} and the numerically integrated 5-loop beta
function \cite{Baikov:2016tgj} with
$\alpha_s(91.187\mathrm{GeV})=0.1181$ \cite{Tanabashi:2018oca}. Then, expression
(\ref{eq:hqe}) for $f^N_b$ and $f^N_t$ becomes:
\begin{equation}
\begin{matrix}
f_b^N&=&0.08748(13) - 0.10129(39) \bar f_4\ ,\\
f_t^N&=&0.09169(4) - 0.09840(10) \bar f_5
\end{matrix}
\end{equation}
where $\bar f_{N_f}$ denotes the sum of scalar, quark contents defined in
(\ref{eq:lnf}).

We input our lattice results into these equations to obtain the final
numbers reported in Table~1. The statistical error originates from the
lattice input while systematic error estimates from the heavy-quark
expansion $(\Lambda_{QCD}/m_b)^2=0.6\%$, from the lattice and from
$\alpha_s$ are combined in quadrature to give the systematic
error. The errors on $f_b^N$ and $f_t^N$ are both dominated entirely by
lattice errors on the light, strange and charm quark contents used as
input quantities.

\section{Arbitrary linear combinations of $\sigma$-terms with full correlations}
\label{sect:tool}

We provide a C routine that computes arbitrary linear combinations of
the scalar, quark contents for the proton, the neutron and the
nucleon, while retaining the full correlation between those
quantities. When called without arguments, it returns the individual
quark contents as well as the Higgs-nucleon coupling, $f^N_h$, and brief
instructions on how to obtain any other linear combination.

The code is available as an ancillary file or via download from
\url{http://particle.uni-wuppertal.de/hch/lincomb.c}

\printbibliography

\end{document}